\documentclass[final,5p,times,twocolumn]{elsarticle}

\usepackage{graphicx}
\usepackage{amssymb}

\usepackage{url}
\usepackage{color} 

\journal{NeuroComputing}

\begin{document}

\begin{frontmatter}


\title{Automatic Segmentation of Gross Target Volume of Nasopharynx Cancer using Ensemble of Multiscale Deep Neural Networks with Spatial Attention}



\author[a]{Haochen Mei}
\author[a]{Wenhui Lei}
\author[a]{Ran Gu}
\author[a]{Shan Ye}
\author[a]{Zhengwentai Sun}
\author[b]{Shichuan Zhang}
\author[a]{Guotai Wang\corref{cor}}
\address[a]{School\ of\  Mechanical\  and\  Electrical \ Engineering, \ University\  of\  Electronic \ Science\  and \ Technology \ of \ China,\  Chengdu,\  China}
\address[b]{Department\ of\ Radiation\ Oncology,\ Sichuan\ Cancer\ Hospital\ and\ Institute,\ University\ of\ Electronic\ Science\ and\ Technology\ of\ China,\ Chengdu,\ China}
\cortext[cor]{Corresponding author: Guotai Wang.}
\ead{guotai.wang@uestc.edu.cn}

\begin{abstract}
\par Radiotherapy is the main treatment method for nasopharynx cancer. Delineation of Gross Target Volume (GTV) from medical images is a prerequisite for radiotherapy. As manual delineation is time-consuming and laborious, automatic segmentation of GTV has a potential to improve the efficiency of this process. This work aims to automatically segment GTV of nasopharynx cancer from Computed Tomography (CT) images. However, it is challenged by the small target region, anisotropic resolution of clinical CT images, and the low contrast between the target region and surrounding soft tissues. To deal with these problems, we propose a 2.5D Convolutional Neural Network (CNN) to handle the different in-plane and through-plane resolutions. We also propose a spatial attention module to enable the network to focus on the small target, and use channel attention to further improve the segmentation performance. Moreover, we use a multi-scale sampling method for training so that the networks can learn features at different scales, which are combined with a multi-model ensemble method to improve the robustness of segmentation results. We also estimate the uncertainty of segmentation results based on our model ensemble, which is of great importance for indicating the reliability of automatic segmentation results for radiotherapy planning. Experiments with 2019 MICCAI StructSeg dataset showed that (1) Our proposed 2.5D network has a better performance on images with anisotropic resolution than the commonly used 3D networks. (2) Our attention mechanism can make the network pay more attention to the small GTV region and improve the segmentation accuracy. (3) The proposed multi-scale model ensemble achieves more robust results, and it can simultaneously obtain uncertainty information that can indicate potential mis-segmentations for better clinical decisions. 
\end{abstract}


\end{frontmatter}


\section{Introduction}

\label{S:1}
Nasopharyngeal Carcinoma (NPC) refers to a malignant tumor that occurs on the walls of nasopharyngeal cavity whose incidence rate is the highest among malignant tumors of the ear, nose and throat. NPC is often found in southern China, Southeast Asia, the Middle East, and the North Africa~\cite{chang2006enigmatic}. According to different conditions of tumor growth, NPC can be divided into four stages. In the T1 phase, the NPC is only at the nasopharynx, and in the T2 phase, the NPC invades parapharyngeal space. If late T3 and T4 phase are reached, the NPC invades the skeletal structure of the skull base and extends to intra-cranial space~\cite{xu2015omission}. Therefore, the earlier the detection and treatment of NPC, the higher the success rate of treatment.

Medical images play a vital role in preoperative decision making, as they provide valuable information about the area, area size, volume, and severity of a tumor. These images can be obtained from different modalities such as Magnetic Resonance Imaging (MRI), Computed Tomography (CT) and Ultrasound, which are all non-invasive techniques. Due to the advantage of imaging speed compared with MRI and higher imaging quality than Ultrasound, CT is the main imaging method for NPC. However, doctors will also use MRI in clinical practice in order to better perform manual annotation.

Radiotherapy is the main treatment for nasopharyngeal carcinoma. Unfortunately, the primary and neck lymphatic drainage areas are easily included in the irradiation field. Therefore, during the actual clinical diagnosis, delineation of Gross Target Volume (GTV) of NPC from medical images such as CT or MRI images is of great importance in radiotherapy planning and follow-up evaluations. However, at present, the delineation task is usually implemented by experienced radiologists through slice-by-slice manual annotation, which is not only tedious, labor intensive and time consuming, but also faced with inter-operator and intra-operator variations. Therefore, automatic delineation methods for GTV region have attracted more and more attention recently~\cite{long2015fully}.
\par
\begin{figure}
\centering\includegraphics[width=1\linewidth]{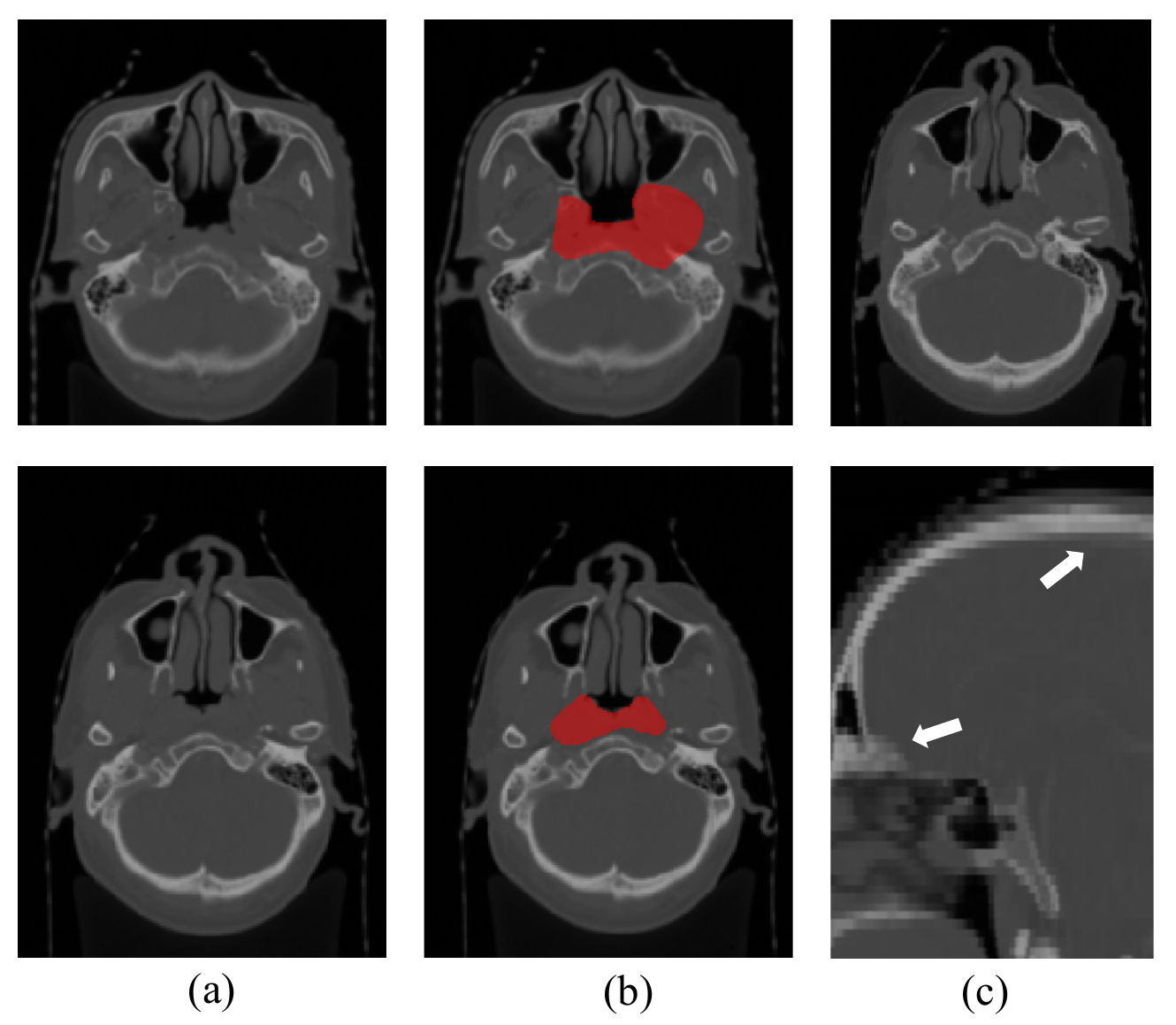}
\caption{The difficulties of GTV segmentation from CT images. (a) Axial CT slices of NPC with low contrast. (b) Manual delineation of GTV in the images from (a). (c) Comparison of the axial and sagittal views of the CT images. Note the high in-plane resolution in the first row and low through-plane resolution marked by white arrows in the second row.}\label{fig:patient}
\end{figure}

 In recent years, methods using deep learning, especially Convolutional Neural Networks (CNN), have been widely applied in medical image analysis. With high-quality and fast automatic segmentation results achieved by Fully Convolution Network (FCN)~\cite{long2015fully}, UNet~\cite{ronneberger2015u}, VNet~\cite{milletari2016v}, and DeepMedic~\cite{kamnitsas2017efficient} etc. , CNNs have been shown to be powerful learning models for segmentation tasks. In spite of that, accurate automatic segmentation for GTV of NPC is still a challenging task. Firstly, most current deep learning-based automatic delineation methods of GTV are implemented on single modality medical images like CT images, which means GTV region has a low contrast with surrounding soft tissues in CT images. Secondly, the boundary of GTV is ambiguous, which makes it difficult to obtain accurate delineation even for human experts. Therefore, it may lead to noisy annotations for training. Thirdly, the NPC only takes up a small region in the whole head and neck CT image, which brings about large imbalance between the segmentation target and the background. In addition, the images are often acquired with high in-plane resolution and low through-plane resolution that results in a large shape change in adjacent slices. Fig.~\ref{fig:patient} illustrates some examples of NPC images showing these challenges.

\par For medical image segmentation tasks, the uncertainty information of segmentation results~\cite{shi2011multi} can help doctors to make better clinical decisions because it can help assess the reliability of segmentation results and identify the challenging situations which would be further reviewed by experienced experts~\cite{jungo2018effect}. In many circumstances even if the model predicts a segmentation result, it cannot be completely believed in the actual clinical diagnosis. For GTV delineation, even human experts can hardly define the exact boundary, so the uncertainty estimation is of great importance for more informed clinical decision. Therefore, for automatic segmentation of GTV, we should not only care about the accuracy of models, but also how confident the model is in the predictions due to the diverse and ambiguous boundaries of GTV region. For example, for areas with high uncertainty like the boundaries and some challenging areas, an experienced expert can be queried for further judgment. For areas with low uncertainty like central regions, the segmentation results tend to be more reliable and experts can pay less attention to that. As a result, radiologists can focus on those indistinguishable areas so that uncertainty can play a guiding role to obtain a better diagnostic result in clinical applications and improve diagnostic efficiency. However, as far as we know, existing deep learning methods for GTV segmentation have rarely analyzed the segmentation uncertainty. 
\par In this paper, we aim at automatic segmentation for GTV of NPC from CT images with anisotropic resolution using novel CNNs, and at the same time we estimate the uncertainty of segmentation results for more informed clinical decisions in radiotherapy planning. The contributions of our method can be summarised as follows: 1) We propose a 2.5D CNN which is designed to better deal with images with high in-plane resolution and low through-plane resolution. 2) We use spatial and channel attention in the network at the same time to improve the performance of CNN in segmentation of the small target like GTV. 3) We propose an ensemble model that combines the predictions of several independent networks trained by global, middle and local scale images to obtain more robust segmentation. 4) Our model naturally leads to an uncertainty estimation of the segmentation results, and we analyze both pixel-level and structure-level uncertainty to better understand the reliability of our segmentation method. Our method was originally designed for the GTV segmentation challenge of 2019 International Conference on Medical Image Computing and Computer-Assisted Intervention\footnote{https://structseg2019.grand-challenge.org/}, and the proposed method won the second place\footnote{http://www.structseg-challenge.org/\#/} among all participants in final competition which recorded 65.29\% in the Dice score and 8.173mm in 95\% HD in the final test set. The code for our method is publicly available\footnote{https://github.com/HiLab-git/Head-Neck-GTV}.

\section{Related Work}
\label{S:2}
\subsection{Automatic Segmentation for Gross Target Volume}
 In 2004, \citet{lee2005segmentation} developed a semi-automatic image segmentation method for NPC that required minimal human intervention and was capable to delineate tumor margins with good accuracy and reproducibility. \citet{han2008atlas} developed a fully automated, atlas-based method for Head and Neck (H\&N) CT image segmentation that employed a novel hierarchical atlas registration approach. Moreover,~\citet{berthon2016atlaas} evaluated the feasibility and impact of a novel advanced auto-segmentation method named ATLAAS in H\&N radiotherapy treatment planning. That method~\cite{berthon2017head} was an automatic decision tree-based learning algorithm for advanced segmentation which was based on supervised machine learning. However, the segmentation accuracy of these methods was limited due to the low contrast and large shape variation of GTV among patients.
 \begin{figure*}
 	\centering\includegraphics[width=1\linewidth]{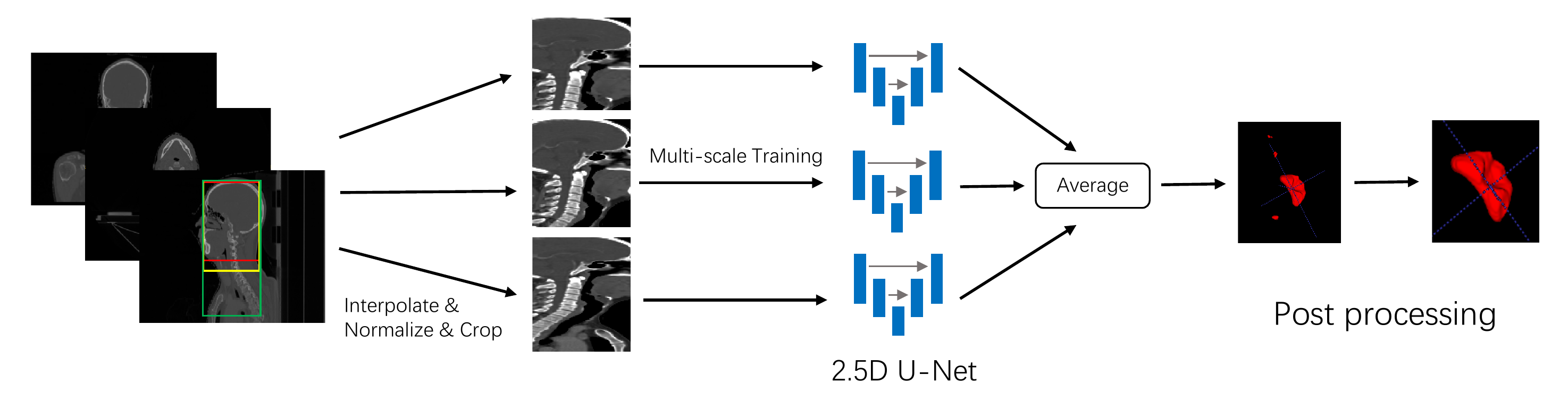}
 	\caption{Overall framework of our proposed method. Probability maps from several separate networks trained by different scales of images are fused to generate the initial segmentation result. Then we take the largest connected area to obtain final segmentation result.}\label{fig:framework}
 \end{figure*}

 \par Applying CNNs to GTV segmentation of nasopharynx cancer has been increasingly studied in recent years. \citet{mohammed2017artificial} exploited neural network to select the right object and input the predicted and extracted texture feature to improve the robustness of segmentation. \citet{men2017deep} developed an end-to-end Deep Deconvolutional Neural Network (DDNN) for GTV segmentation which enabled fast training and testing.~\citet{ma2017automatic} trained three deep single-view CNNs separately to improve the accuracy of NPC segmentation. 
Recently, an automatic GTV segmentation framework based on deep dense multi-modality network was proposed by \citet{guo2019gross} to deal with the low background contrast and potential artifacts in conventional planning of CT images. In 2018, a multi-modality MRI fusion network (MMFNet)~\cite{chen2018mmfnet} based on three modalities of MRI (T1, T2 and contrast-enhanced T1) was proposed to achieve accurate segmentation of NPC. However, despite these proposed methods, accurate segmentation of GTV of NPC from CT images still remains a challenging task.

\subsection{Attention for Segmentation}
\par Attention mechanism has been increasingly used for deep learning which enables the model to focus more on the most important part of feature maps for better segmentation performance.~\citet{oktay2018attention} proposed Attention Gate (AG) using deep feature maps to recalibrate the region of intrest.~\citet{hu2018squeeze} proposed a `Squeeze and Excitation' network using channel attention to emphasize the relevant feature channels and suppress the irrelative parts.~\citet{woo2018cbam} proposed Convolutional Block Attention Module (CBAM) which extracted feature information by mixing cross-channel and spatial information.~\citet{roy2018concurrent} squeezed and excited feature maps in its spatial-wise and channel-wise manners for more accurate segmentation of small targets.
\citet{wang2018non} proposed non-local operations in order to capture long-range dependencies for better segmentation. A dual attention network (DANet) was proposed in \cite{fu2019dual} to combine local features with global dependencies. There are also various forms of attentions leveraged for semantic segmentation~\cite{li2018pyramid,chen2016attention,huang2019ccnet,li2018tell}. For  3D attention, \citet{rickmann2019project} extended channel-wise attention mechanism into three dimentions through `Project and Excite' operation. However, to the best of our knowledge, the effect of these mechanisms on GTV segmentation has rarely been investigated. 

\subsection{Segmentation Uncertainty}
 Extensive researches have been investigated on different existing medical image segmentation tasks by using uncertainty estimation. Shape and appearance prior information were used by \citet{saad2010exploration} to estimate the segmentation uncertainty.~\citet{shi2011multi} explored how to use uncertainty to improve the robustness of graph cut-based cardiac image segmentation system.~\citet{prassni2010uncertainty} visualized the uncertainty of a random walker-based segmentation to guide volume segmentation of brain MRI and CT.~\citet{sankaran2015fast} estimated lumen segmentation uncertainty for realistic patient-specific blood flow modeling.

\par For deep CNNs, both epistemic and aleatoric uncertainty have been investigated in recent years. For model (epistemic) uncertainty, Bayesian networks provide a mathematical-based calculation method, but it is difficult to implement and consumes a lot of computing resources. Alternatively, it was shown that dropout at test time could be considered as a Bayesian approximation to estimate model uncertainty~\cite{li2017compactness}. Stochastic Variational Gradient Descent (SVGD) was also adopted by \citet{zhu2018bayesian} to approximate Bayesian inference for uncertain CNN parameters. 

\citet{lakshminarayanan2017simple} proposed the method of ensembling multiple models for uncertainty estimation, which was simple to implement. For test image-based (aleatoric) uncertainty, Kendall and Gal~\cite{kendall2017uncertainties} introduced a Bayesian deep learning framework to obtain aleatoric uncertainty from input data and combined it with epistemic uncertainty. \citet{wang2019aleatoric} proposed aleatoric uncertainty estimation with test-time augmentation for medical image segmentation. Based on these studies, uncertainty can play a good guiding role in medical image segmentation for radiologists. But as far as we know, hardly anyone has performed uncertainty analysis on the  GTV segmentation task of nasopharynx cancer. 

\section{Methodology}
\begin{figure*}[h]
	\centering\includegraphics[width=1\linewidth]{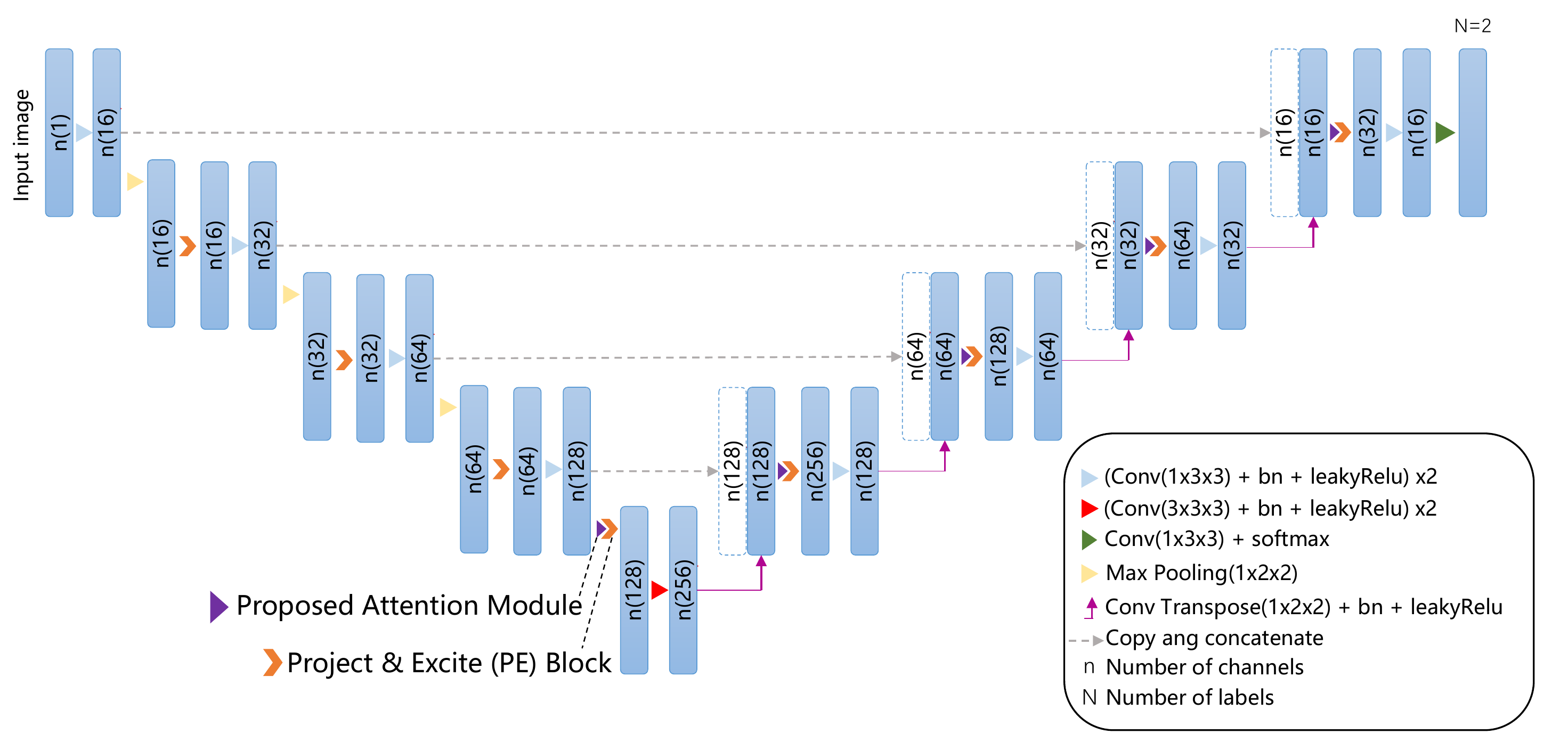}
	\caption{Proposed 2.5D network architecture for GTV segmentation from CT images with large inter-plane spacing. Dotted and solid boxes indicate copied data and operation output. The number n in each block represents channel numbers. We use 2D convolutions($1\times3\times3$) in both encoder and decoder convolution blocks while only use 3D convolutions($3\times3\times3$) in the bottom block. }
	\label{fig:network}
\end{figure*}
Our method consists of four main parts: 1) Data processing based on the truncation of HU, intensity normalization and image cropping; 2) A 2.5D CNN with a combination of in-plane attention module and Project \& Excite (PE) block~\cite{rickmann2019project} for gross target volume segmentation; 3) Using a model ensemble method based on multi-scale information fusion. 4) Our model ensemble method naturally leads to an uncertainty estimation of segmentation results, which can indicate potential mis-segmentations for better clinical decision making. Fig.~\ref{fig:framework} shows the overall framework we propose.

\subsection{Data and Preprocessing}
We used the dataset of MICCAI 2019 StructSeg challenge (GTV segmentation task) for experiments. The dataset consists of CT images of 50 NPC patients with high in-plane resolution around $1\times1$mm and in-plane size $512\times512$. The inter-slice spacing is 3mm and the slice number is in the range of 103 to 152. In the end, the MICCAI 2019 StructSeg Challenge used 10 unpublished data for final testing. The ground truth for training set was manually annotated by experienced neurosurgeons and physicists. Because the official test set was not publicly available, we randomly split the official training images into 40 and 10 for training and testing respectively. 
\par For data preprocessing, we first truncate the intensity values of all images to the range of [-200, 700] HU to increase the contrast of the target area and then normalize it by the mean value and mean standard deviation. Furthermore, to maintain the same resolution, the pixel spacing of all images in the x, y, and z directions is uniformly interpolated to $1\times1\times3$ mm in order to get better training models. 

\begin{figure}
	\centering\includegraphics[width=1\linewidth]{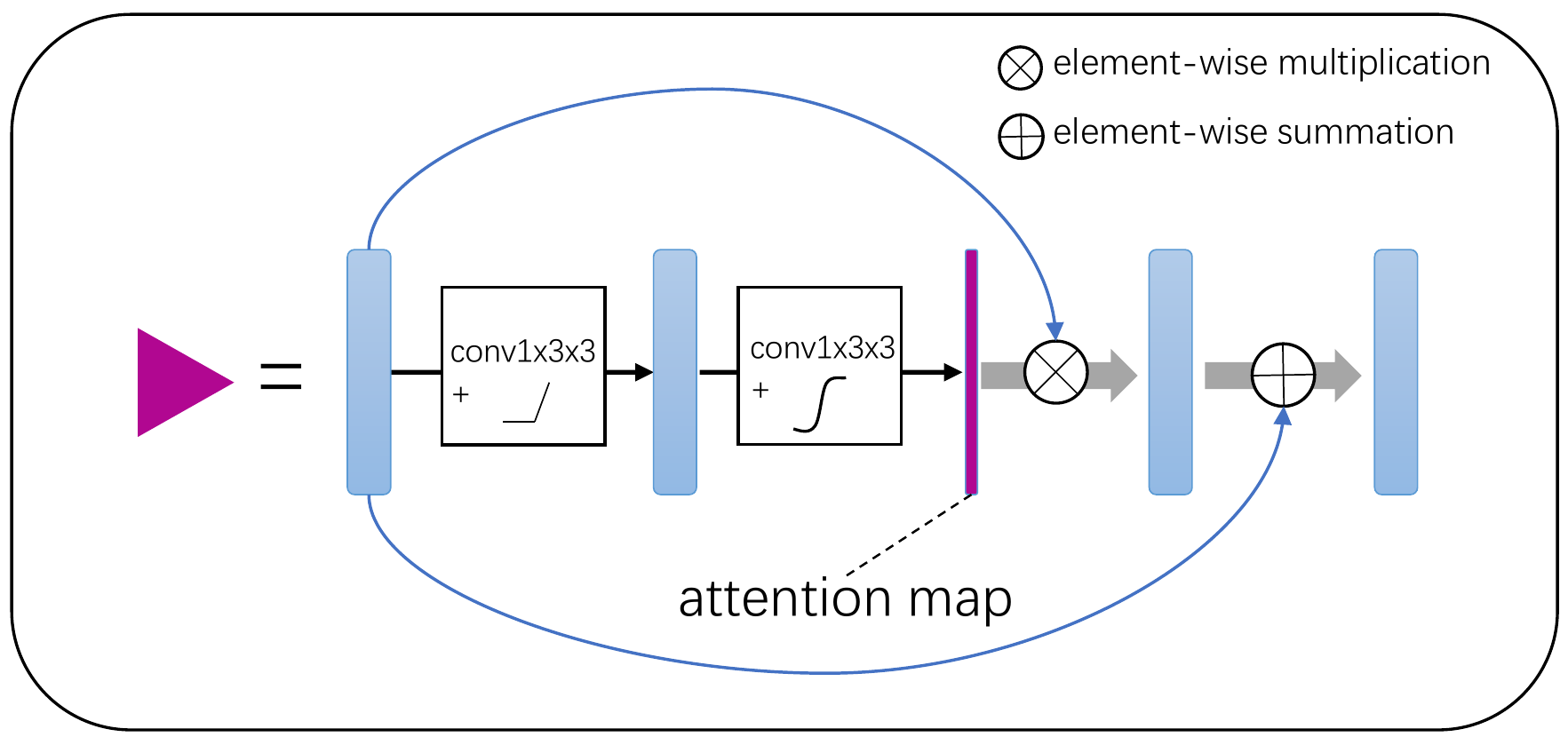}
	\caption{The proposed Attention Module (AM) with two $1\times3\times3$ convolutions produces an attention map to recalibrate the feature map.}
	\label{fig:attention}
\end{figure}

\begin{figure*}
	\centering\includegraphics[width=1\linewidth]{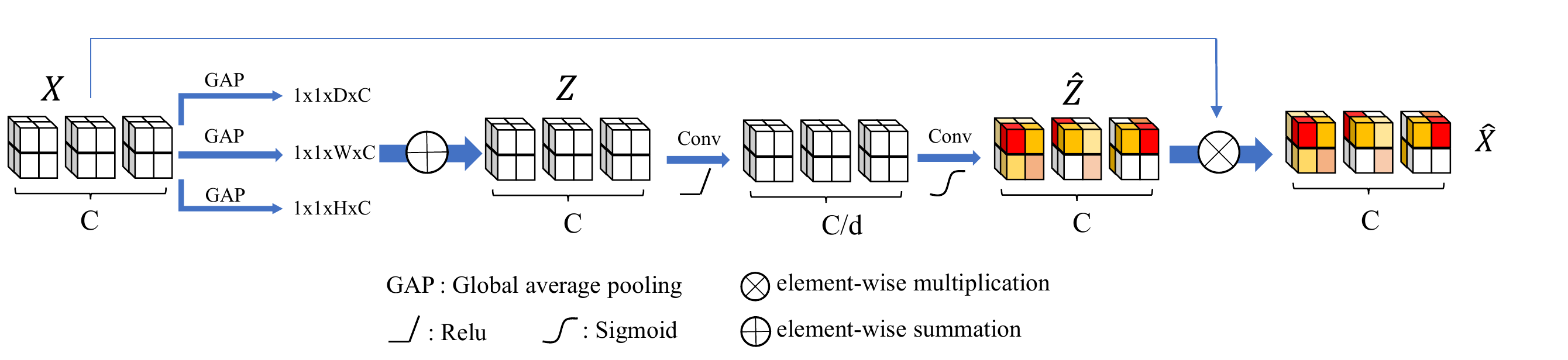}
	\caption{The structure of PE block. The block learns the relevant information of space and channel by performing Projection operation with three different pooling layers and Excitation operation with two convolution layers.}
	\label{fig:peblock}
\end{figure*}
\subsection{2.5D Network with In-plane Spatial Attention and Channel Attention}
\par{\bfseries Overall Network Architecture.} Fig.~\ref{fig:network} shows the overall network architecture we use. The main structure of our network follows the typical encoder and decoder design of UNet~\cite{ronneberger2015u}. As shown in the Fig.~\ref{fig:network}, a total of nine convolution blocks are used and each of them contains two convolution layers followed by Batch Normalization (BN) and Leaky Rectified Linear Units (leaky ReLU). And each convolution block is preceded by a PE block~\cite{rickmann2019project} except the first one because the input of which is input image with one channel. The attention module is placed before PE block in the decoder and the bottom block to capture spatial information of small GTV region. The main purpose of PE block is to obtain channel-wise information and also take into account spatial information. We only perform three downsamplings and the deconvolution layers are adopted to upsample the feature maps. The final layer is composed of a convolution layer and the softmax function which provides the segmentation probabilities.
\par{\bfseries 2.5D CNN.} Standard 2D CNNs cannot take into account the correlation between slices, which will reduce the overall segmentation performance when dealing with 3D targets. For typical isotropic 3D CNNs, it is generally necessary to upsample the resolution along three axes of image to a uniform value to balance the physical receptive field along each axis, which usually leads to higher memory requirements and may limit the depth of CNNs. At the same time, because of the low through-plane resolution and the distinct contrast between the bone and nearby soft tissues, resampling images to isotropic resolution will produce plenty of artifacts on interpolated slices which may mislead the segmentation results. Therefore, we combine the $3\times3\times3$ convolutions with $1\times3\times3$ convolutions due to the large difference in in-plane resolution and through-plane resolution to design a 2.5D CNN, where we use 3D convolutions ($3\times3\times3$) at the bottleneck of the encoder-decoder structure and 2D convolutions in other blocks, as shown in Fig.~\ref{fig:network}. Note that this is different from existing 2.5D networks that refer to taking several adjacent slices as input and predicting the middle slice \cite{vu2019evaluation} or applying 2D CNNs in axial, sagittal and coronal views respectively \cite{prasoon2013deep}. For our 2.5D network, both the input patch size and output patch size were $16\times64\times64$, i.e., they have the same slice number.

\par{\bfseries In-plane Attention Module.} In the GTV segmentation task, it is still difficult to improve the segmentation accuracy of small objects with large shape changes. Since the GTV region and the surrounding soft tissues have similar intensity values, it is desirable to enable the network to better focus on the feature information of GTV region in a large image context. Many previous works~\cite{oktay2018attention} have shown that spatial attention helps to enable the network to focus on learning the target area throughout the whole image context. Therefore, in this paper we also propose an attention module to improve the segmentation accuracy of GTV region, which can make good use of spatial information so that the network can focus on the target region. As shown in Fig.~\ref{fig:attention}, the attention module consists of two convolution layers with kernel size of $1\times3\times3$. The first convolution layer reduces the number of channels to half, and it is followed by a ReLU activation function. The second convolution layer further reduces the number of channels to 1, and then generates a spatial attention map through the sigmoid activation function. The spatial attention map is a single-channel feature map of attention coeffcient ${\alpha_i}\in[0,1]$ which indicates the relative importance for each spatial position $i$. The spatial attention map is then multiplied by the input feature map. Moreover, a residual connection is used in our attention module for better convergence.
\par{\bfseries Project \& Excite (PE) Block.} The 3D cSE~\cite{zhu2019anatomynet} module squeezes spatial information of 3D volume feature map into one scalar value corresponding to each channel which leads to the loss of spatial information. For segmentation of small targets such as the GTV of NPC, such spatial information is important for final segmentation result. Compared with 3D cSE block, the PE block~\cite{rickmann2019project} can retain the spatial information of feature maps through projection operation. Considering a feature map with shape $D\times H\times W\times C$, where D, H, W, C are the depth, height, width and channel number respectively. Our attention module squeezes the feature map along the channel dimension and obtain a spatial attention map with shape $D\times H\times W$, i.e., obtain the attention coefficient (a scalar) for each spatial position. However, the PE block first projects the feature map along each axis to obtain three feature maps with shapes $D\times 1\times 1\times C$, $1\times H\times 1\times C$ and $D\times H\times W\times C$ respectively, then expands them to the original shape and adds them together, and finally it obtains an attention map with shape $D\times H\times W\times C$. It does not provide a single scalar to indicate the importance of each spatial position. Therefore, PE block is more related to channel attention and it assigns a voxel-wise attention coefficient to each channel. Due to these differences, PE block and our explicit spatial attention module are complementary to each other. As shown in Fig.~\ref{fig:peblock}, the `Project \& Excite' module consists of two parts including the projection and excitation operations. The projection uses an average pooling operation for each dimension, which can remain more spatial information than the spatial squeezing operation. The excitation learns inter-dependencies between projections of different channels. Therefore, the PE block is capable to combine spatial and channel context for recalibration. The architectural details of PE block is illustrated in Fig.~\ref{fig:peblock}.
\par For projection, let $x_c$ represents the $c-th$ channel of input feature map $X$ and $z_{h_c}$, $ z_{d_c}$, $z_{d_c}$ denote the output of average pooling operation for each dimension respectively. The detailed definition is shown as follows:
\begin{equation}
\label{eq:emc}
    z_{h_c}\left(i\right) =\frac{1}{W}\frac{1}{D}\sum_{j=1}^{W}\sum_{k=1}^{D}x_c\left(i,j,k\right), i \in \{1,...,H\}
\end{equation}

\begin{equation}
\label{eq:emc}
    z_{w_c}\left(j\right) =\frac{1}{H}\frac{1}{D}\sum_{i=1}^{H}\sum_{k=1}^{D}x_c\left(i,j,k\right), j \in \{1,...,W\}
\end{equation}

\begin{equation}
\label{eq:emc2}
    z_{d_c}\left(k\right) =\frac{1}{H}\frac{1}{W}\sum_{i=1}^{H}\sum_{j=1}^{W}x_c\left(i,j,k\right), k \in \{1,...,D\}
\end{equation}
Then each of $z_{h_c}$, $ z_{d_c}$, $z_{d_c}$ is spreaded to the original shape of input feature map $X$ which is $H\times W\times D\times C$. Then these spreaded tensors are added to get $Z$ as the input of excitation operation $F_{ex}$. The details of excitation operation is as follows:
\begin{equation}
\label{eq:emc3}
    F_{ex}\left(Z\right) = \varrho\left(F_{2}\left(\varphi \left(F_{1}\left(Z\right)\right)\right)\right)
\end{equation}
\par For excitation, $F_{1}$ and $F_{2}$ denote the convolution layers followed by ReLU $\varphi$ and sigmoid $\varrho$ activation functions respectively. $F_{1}$ reduces the number of channels to $C/d$. Then $F_{2}$ recovers the channel number to its original number. The final output $\hat{X}$ of PE block is obtained by an element-wise multiplication of $X$ and $\hat{Z}$ representing the output of $F_{ex}$. Detailed definition can be seen as follows:
\begin{equation}
\label{eq:emc4}
    \hat{X} = X\odot\hat{Z} = X\odot\left(F_{ex}\left(Z\right)\right)
\end{equation}

\subsection{Multi-Scale Sampling}
The GTV region in our segmentation task is very small in the full CT image context. Training a network using entire image as input can make use of global context but is faced with large imbalance between the foreground and the background. Moreover, training the network with local patches around the target can alleviate the imbalance problem but easily leads to many false positives at test time. To tackle these problems, we propose a multi-scale sampling method. Specifically, we crop the images for performing multi-scale sampling to obtain patches for training of CNNs, as shown in Fig.~\ref{fig:framework}. In the x,y direction, we crop the images based on a rough bounding box of the head region. For local sampling, we sample the patches only form the region of head. The middle sampling strategy samples patches from a larger region including both the head and the neck. As for global sampling, it obtains patches from the entire image region. During the training process, we use these three sampling strategies to train segmentation models respectively. Due to the different sampling strategies, these models are able to use features at different scales for segmentation, and their results will be fused for more robust segmentation, as described in the following section. 

\subsection{Model Ensemble}
Ensemble is an effective way to further improve performance which has been used by many participants in previous segmentation challenges~\cite{bakas2018identifying}. We also employ model ensemble to improve the robustness of segmentation results. For each sampling strategy, we train two models, and finally we obtain six models for ensemble. Then we perform an averaging operation on the output probability maps of six CNNs. We take argmax operation on the mean probability map to get initial segmentation result. Finally we take the largest connected region for postprocessing to get final segmentation result which can reduce false positives.

\subsection{Segmentation Uncertainty}
\par Uncertainty is typically estimated by measuring the diversity of predictions for a given image~\cite{kendall2017uncertainties}. Let $Y$ represent the discretized labels obtained by the argmax operation at the last layer of the network and $X$ denote the input images. Using the variance and entropy of the distribution $p(Y|X)$ are two common methods for uncertainty estimation. Since our models have multiple predictions for ensemble, these predictions naturally lead to an uncertainty estimation, where a higher diversity among these predictions indicates a higher uncertainty in segmentation results. Suppose $\mathcal{Y}^i$ denotes the predicted label for the $i-th$ pixel. With the multiple model predictions, a series of values of 
$\mathcal{Y}^i=\{y_1^i,y_2^i,y_3^i,...,y_N^i\}$ can be obtained. Let $\hat{p}_n^i$ represents the frequency of $n-th$ unique value in $\mathcal{Y}^i$. Following ~\cite{wang2019aleatoric}, we define the pixel-wise uncertainty based on the entropy information: 
\begin{equation}
\label{eq:emc5}
    H\left(Y|X\right) \approx -\sum_{n=1}^{N}\hat{p}_n^iln\left(\hat{p}_n^i\right)
\end{equation}

\par For $n$ prediction samples from different models, let $V=\{v_1,v_2,v_3,...,v_n\}$ represent the volume collection, where $v_i$ is the volume obtained by the $i-th$ model. Assume $\sigma_v$ and $\mu_v$ represent the standard deviation and mean value of $V$ respectively. We use the Volume Variation Coefficients (VVC) to estimate the structure-wise uncertainty which is defined as follows: 
\begin{equation}
\label{eq:emc6}
    VVC =\frac{\sigma_v}{\mu_v}
\end{equation}

\section{Experiments and Results}
\subsection{Implementation and Evaluation Methods}
The proposed framework was implemented in PyTorch with two NVIDIA GTX 1080 Ti GPUs. For training, the Adam optimizer was adopted with weight decay $1 \times 10^{-5}$, batch size 16. The learning rate was initialized to $1 \times 10^{-4}$, and was decayed by 0.9 every 10k iterations. The networks were trained with the Dice loss function~\cite{milletari2016v}. Random cropping and random flipping were used for data augmentation. Quantitative evaluations of segmentation accuracy are based on Dice score, Average Symmetric Surface Distance (ASSD) and Relative Volume Error (RVE).

\begin{equation}
\label{eq:emc7}
Dice =\frac{2 \times TP}{2\times TP + FN + FP}
\end{equation}
where TP, FP and FN are true positive, false positive and false negative respectively. And the definition of ASSD is:
\begin{equation}
\label{eq:emc8}
ASSD =\frac{1}{\left|S\right| + \left|G\right|}\left( \sum_{s\in S}d\left(s,G\right) + \sum_{g\in G}d\left(g,S\right)\right)
\end{equation}
where $S$ and $G$ denote the set of surface points of a segmentation result and the ground truth respectively. $d(s,G)$ is the shortest Euclidean distance between a point $s\in S$and all the points in $G$. RVE can be calculated by:

\begin{equation}
\label{eq:emc9}
RVE =\left|\frac{V_{pre} - V_{gt}}{V_{gt}}\right| \times 100\%
\end{equation}
where $V_{pre}$ and $V_{gt}$ denote the volume of predicted segmentation results and the volume of ground truth respectively. 
\begin{figure}
	\centering\includegraphics[width=1\linewidth]{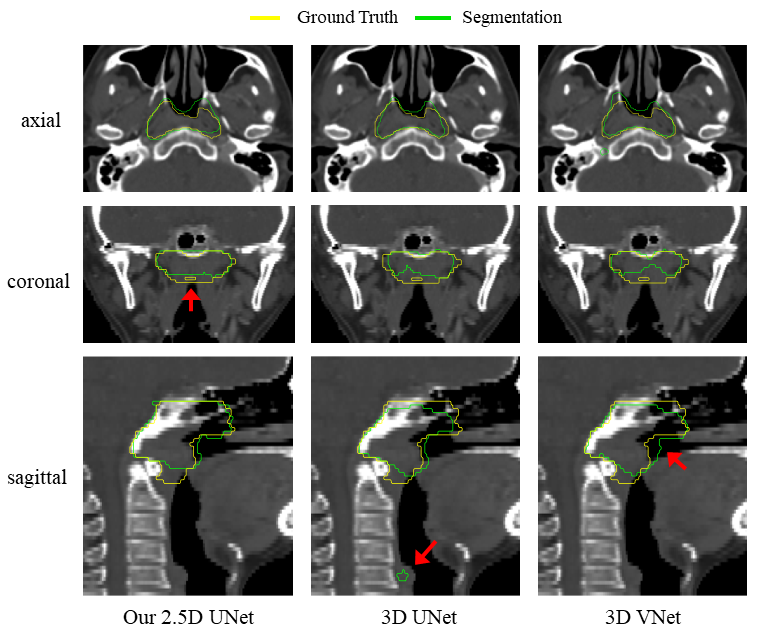}
	\caption{Visual comparison of different backbone networks for GTV segmentation. The red arrows in the second and third column denote the false positives and poorly segmented areas of VNet respectively.
	}
	\label{fig:backbone_compare}
\end{figure}

\subsection{Comparison of Different Networks}
First of all, we evaluated the segmentation performance of different backbone networks: 3D UNet~\cite{ronneberger2015u}, 3D VNet~\cite{milletari2016v}, the proposed 2.5D UNet without PE block~\cite{rickmann2019project} and attention block. It is noted that in this setting, this 2.5D UNet and 3D UNet follow the same structure except the kernel size. And all these networks were trained with local sampling strategy. Table 1 shows the performance of different backbones under three different evaluation criteria. We can observe that the 2.5D UNet has better performance in terms of Dice score, ASSD and RVE than 3D UNet and 3D VNet, which verifies the effectiveness of our proposed backbone. In terms of inference time, our network is faster than 3D VNet and 3D UNet. The visual segmentation results of three backbone networks are shown in Fig.~\ref{fig:backbone_compare}. From the figure we can see that compared with 3D UNet and 3D VNet, our proposed 2.5D UNet produces fewer false positive predictions. While from the segmentation results of coronal view and sagittal view, we can clearly see that the prediction of our proposed 2.5D network has more robust segmentation results. In short, the experimental results show that our 2.5D backbone network has better performance than its 3D counterpart or variant when dealing with images with anisotropic 3D resolution.

\begin{table}
\scalebox{0.95}{\centering
\begin{tabular}{c c c c c}
\hline
\textbf{Network} & \textbf{Dice(\%)} & \textbf{ASSD(mm)} & \textbf{RVE(\%)} & \textbf{Time(s)}\\
\hline
3D UNet & 59.91 & 5.52 & 72.7 & 0.094\\
3D VNet & 61.02 & 5.63 & 65.0 & 0.123\\
2.5D UNet & \textbf{62.16} & \textbf{5.04} & \textbf{60.6} & \textbf{0.093}\\
\hline
\end{tabular}}
\caption{Quantitative evaluations of different backbones for GTV segmentation including Dice, ASSD, RVE and inference time. Additionally, the results of inference time are tested with local sampling strategy.}
\end{table}

\subsection{Effect of Different Modules}
\par We further explored the effect of PE block and our proposed Attention Module (AM). We combined these modules with our 2.5D backbone respectively. We used eight PE blocks and five attention modules, as shown in Fig.~\ref{fig:network}. We compared four variants: our proposed 2.5D UNet described above, 2.5D UNet using PE block, 2.5D UNet using PE block and AM and 2.5D UNet using PE block and AG~\cite{oktay2018attention}. All these variants were trained using local sampling strategy. It can be seen from Table 2 that after adding PE block and the proposed AM, multiple evaluation indicators in terms of the Dice, ASSD and RVE are all improved, which proves that both modules can improve the performance of network. In addition, our AM is more efficient and has a better segmentation performance compared with AG.

\begin{table}
\scalebox{0.8}{\centering
\begin{tabular}{c c c c c}
\hline
\textbf{Network} & \textbf{Dice(\%)} & \textbf{ASSD(mm)} & \textbf{RVE(\%)} & \textbf{Time(s)}\\
\hline
2.5D UNet & 62.16 & 5.04 & 60.6 & \textbf{0.093} \\
2.5D UNet + PE & 63.11 & 4.35 & 55.0 & 0.097 \\
2.5D UNet + PE + AG & 62.62 & 4.41 & 50.5 & 0.099  \\
2.5D UNet + PE + AM & \textbf{64.46} & \textbf{4.24} & \textbf{50.0} & 0.098 \\
\hline
\end{tabular}}
\caption{Comparison of the effects of different modules superimposed with 2.5D UNet. AM: Our proposed attention module. AG : The Attention Gate proposed in~\cite{oktay2018attention}. PE: The PE block proposed in~\cite{rickmann2019project}.}
\end{table}

\begin{figure*}[h]
	\centering\includegraphics[width=0.7\linewidth]{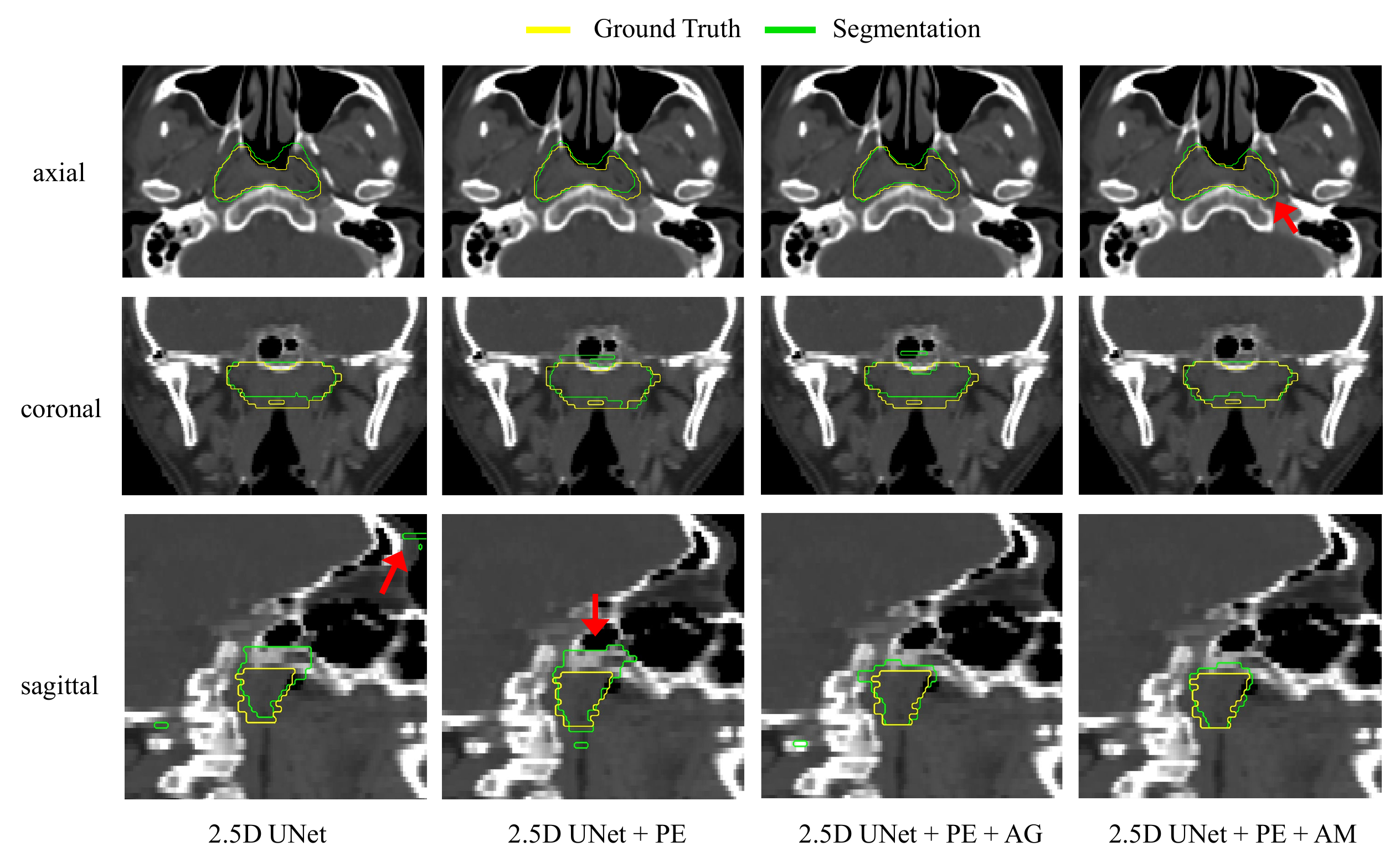}
	\caption{Visual comparison of different variants of 2.5D UNet. The red arrows in the first, second and fourth column denote the false positive regions, the areas with poor and good segmentation result respectively. In the figure, green curves and yellow curves denote the segmentation results and ground truth respectively.}
	\label{fig:module_compare}
\end{figure*}
\begin{figure*}[h]
	\centering\includegraphics[width=0.85\linewidth]{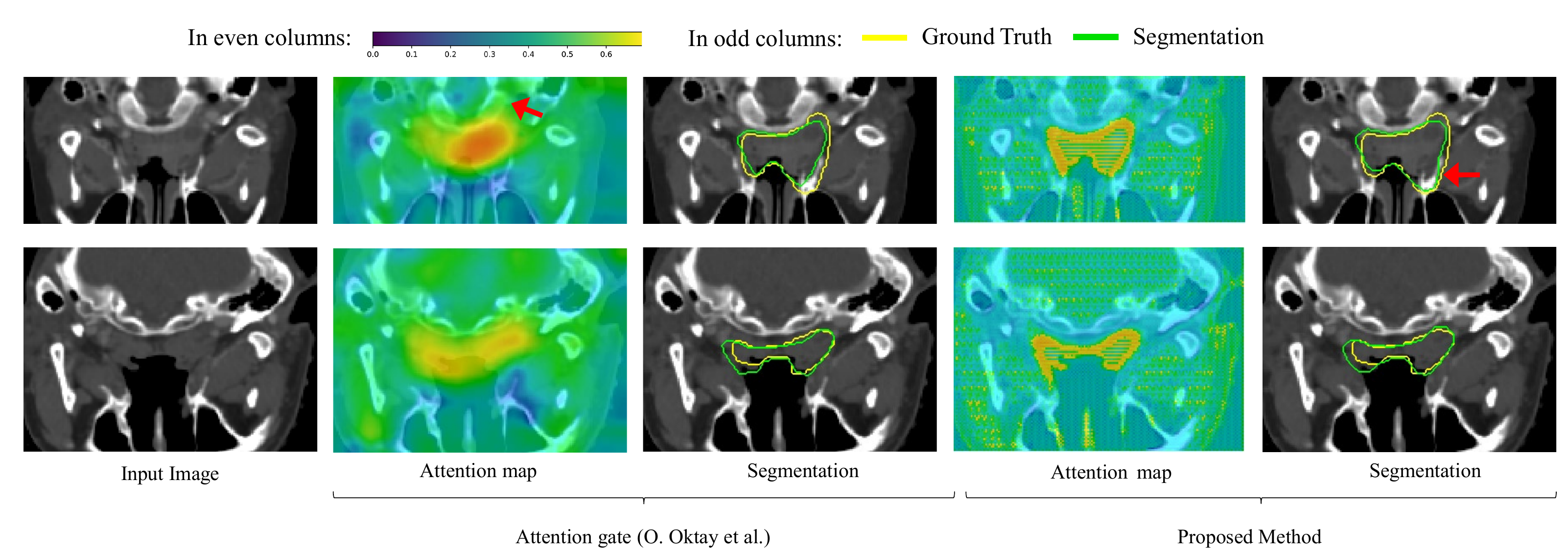}
	\caption{Visual comparison of attention maps obtained by Attention Gate (AG) and our proposed Attention Module (AM). Odd columns: original images, their segmentation results and ground truth. Even columns: attention maps from AG~\cite{oktay2018attention} and our AM, where warmer color represents higher attention. The red arrows denote some unrelated areas noticed by AG and a better boundary segmentation result of our AM respectively. Note that two rows are form different patients.}
	\label{fig:attention_compare}
\end{figure*}
\par Fig.~\ref{fig:module_compare} shows a visual comparison of three different modules. It can be observed that PE block and AM can efficiently improve the segmentation performance, especially in the boundaries of GTV. As we can see in the figure, 2.5D UNet + PE + AM outperforms other networks especially in the axial and sagittal views. It not only produces fewer false positives but also has better segmentation performance of boundaries. Moreover, Fig.~\ref{fig:attention_compare} shows the visualization of attention maps obtained by AG and our proposed AM. Though the attention map of AG successfully suppresses most part of the background region, it can be seen from the visualization that some unrelated areas are still noticed just as shown by the red arrow in the second column. In addition, AG does not pay much attention to the boundaries of GTV region while our AM can concentrate more on the boundaries, which helps to produce a better boundary segmentation effect as indicated by the red arrow in the last column of Fig.~\ref{fig:attention_compare}. In general, the proposed AM does not only highlight the GTV region, but also concentrate more on the boundary of GTV area and the overall structural information which leads to better segmentation of GTV.

\subsection{Effect of Multi-Scale Model Ensemble}
\begin{figure*}
\centering\includegraphics[width=0.75\linewidth]{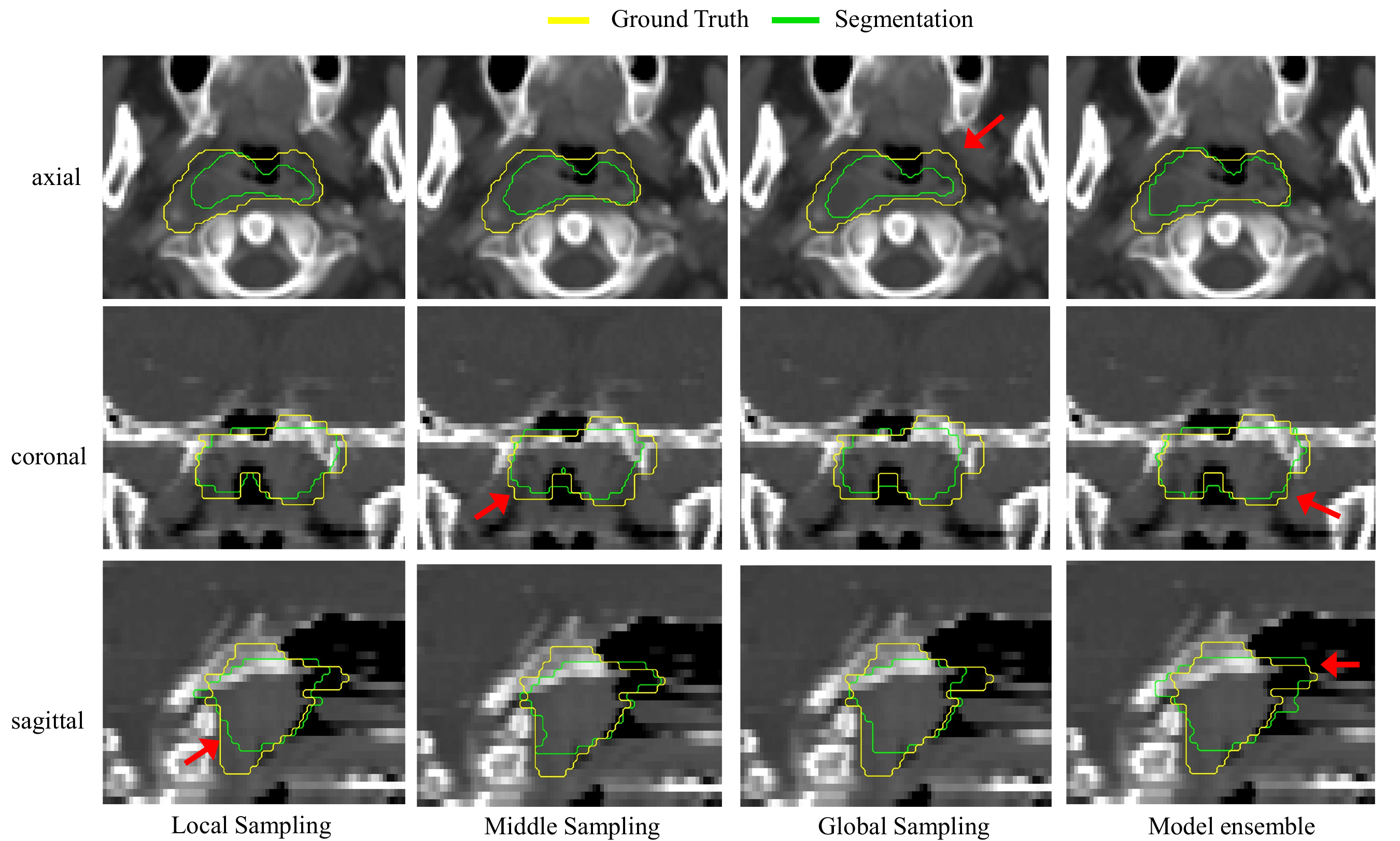}
\caption{Visual comparison between a single model trained with different sampling strategies and model ensemble. The red arrows in the first three columns represent some bad segmentation results and the red arrows in the last column represent better segmentation results obtained by our model ensemble.}
\label{fig:ensemble_compare}
\end{figure*}

As we mentioned before, sampling at a single scale during training would cause false positives or imbalance between the target area and the background area. In our work, the final GTV segmentation results were obtained by averaging the probability map from six different models with the structure of 2.5D UNet + PE + AM in order to fuse the feature information of three different scales. From Table 3 it can be seen that after the ensemble of models obtained by multi-scale sampling training, the model performs better under three indicators including Dice, ASSD and RVE, which means the model is more robust in segmentation of GTV. A single model takes around 0.1s and the ensemble takes less than 0.3s for inference. The segmentation results of model ensemble are visualized in Fig.~\ref{fig:ensemble_compare}. We can learn from the figure that global sampling usually results in under-segmentation while local and middle sampling usually bring about some over-segmentation. Through the model ensemble of three scales, we can get better segmentation results on the boundaries of GTV.

\begin{table}
\scalebox{0.88}{\centering
\begin{tabular}{c c c c c}
\hline
\textbf{Network} & \textbf{Dice(\%)} & \textbf{ASSD(mm)} & \textbf{RVE(\%)} & \textbf{Time(s)}\\
\hline
Local sampling & 64.46 & 4.24 & 50.0 & \textbf{0.098} \\
Middle sampling & 64.69 & 4.15 & 59.2 & 0.099 \\
Global sampling & 64.54 & 4.08 & 41.5 & 0.10  \\
Model Ensemble & \textbf{65.66} & \textbf{3.98} & \textbf{40.4} & 0.297\\
\hline
\end{tabular}}
\caption{Comparison of different sampling methods and model ensemble.}
\end{table}

\subsection{Segmentation Results with Uncertainty}
\begin{figure*}
	\centering\includegraphics[width=0.8\linewidth]{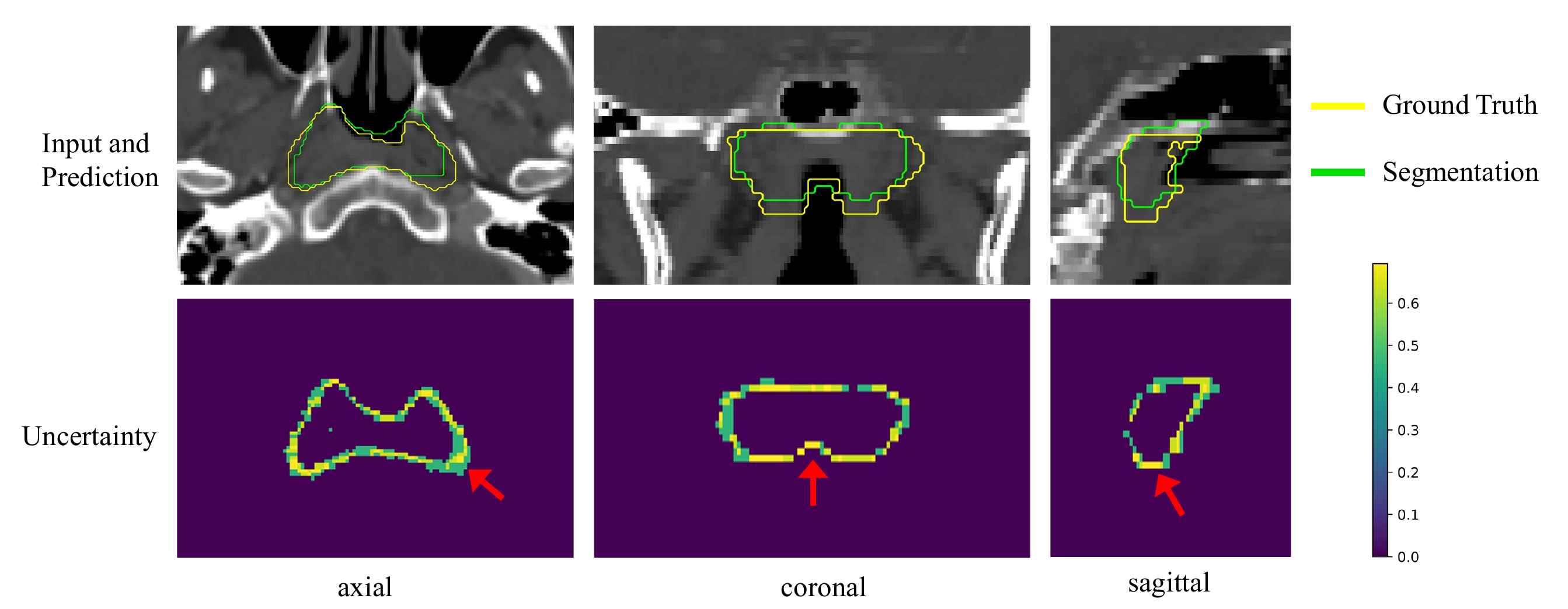}
	\caption{Segmentation results and pixel-wise uncertainty information based on our model ensemble. The purple pixels correspond to lower uncertainty values while the yellow pixels correspond to higher uncertainty values.}
	\label{fig:uncertainty_visual}
\end{figure*}

Fig.~\ref{fig:uncertainty_visual} shows the visualization of pixel-wise uncertainty obtained by our model ensemble method with CNNs trained at multiple scales. The first row shows the input image and its segmentation obtained by a single 2.5D UNet + PE + AM. The bottom row visualizes the pixel-wise uncertainty. In the uncertainty map, the purple pixels have lower uncertainty values while the yellow pixels have higher uncertainty values. The uncertainty map is represented by the entropy of $N$ predicted pixels. As shown in the figure, it shows an indeterminate segmentation not only on the boundaries of GTV but also in some areas that are difficult to segment, as indicated by the red arrows in the figure. As an example, the red arrow in the uncertainty map of axial view indicates the region of high uncertainty value whose predictions of corresponding region in input image is an under-segmentation.

\subsection{Correlation between Uncertainty and Segmentation Error}

\begin{figure}[h]
	\centering\includegraphics[width=0.8\linewidth]{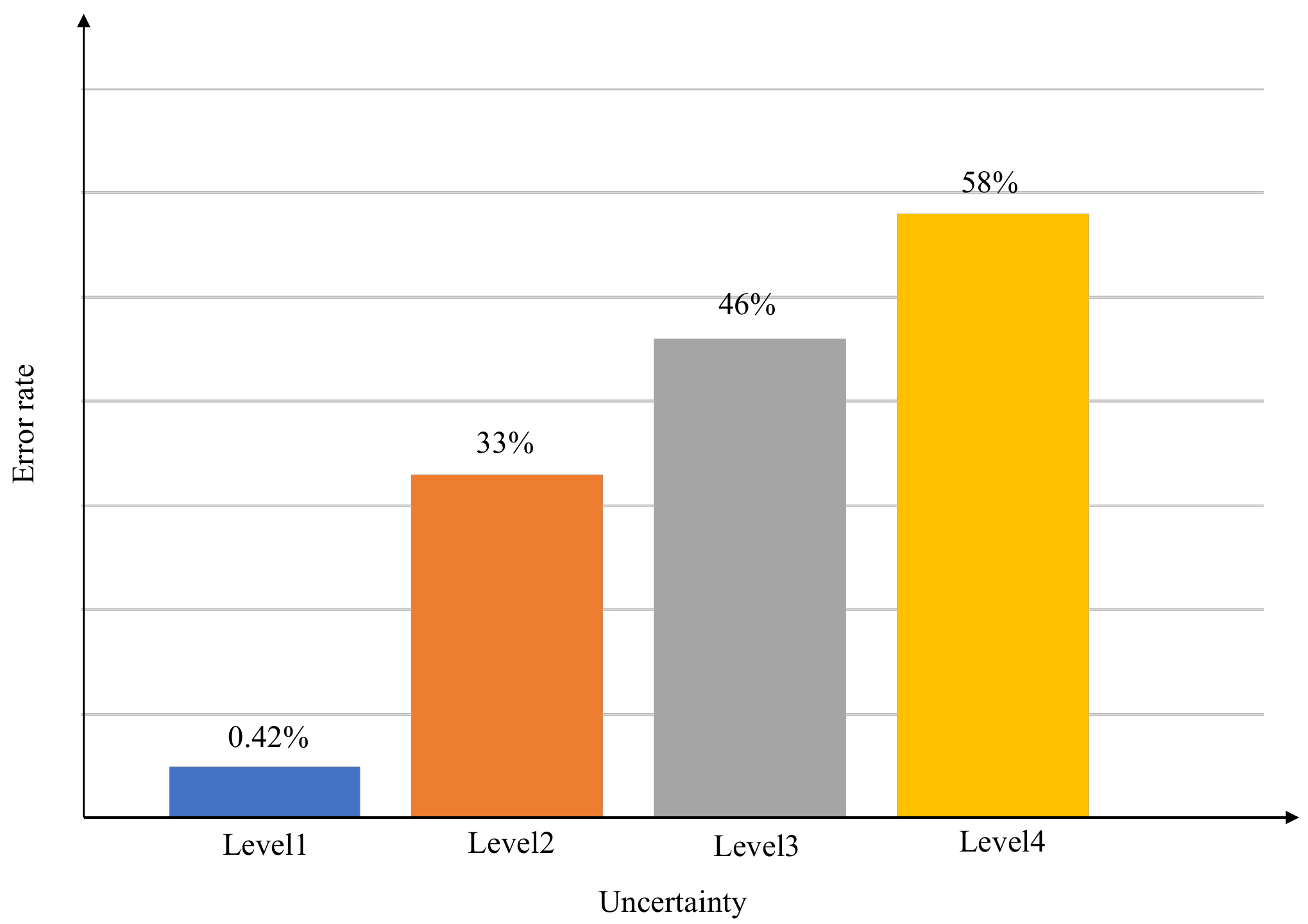}
	\caption{Pixel-level prediction error rate at different uncertainty levels.}
	\label{fig:pixel_error}
\end{figure}

\begin{figure}
	\centering\includegraphics[width=0.8\linewidth]{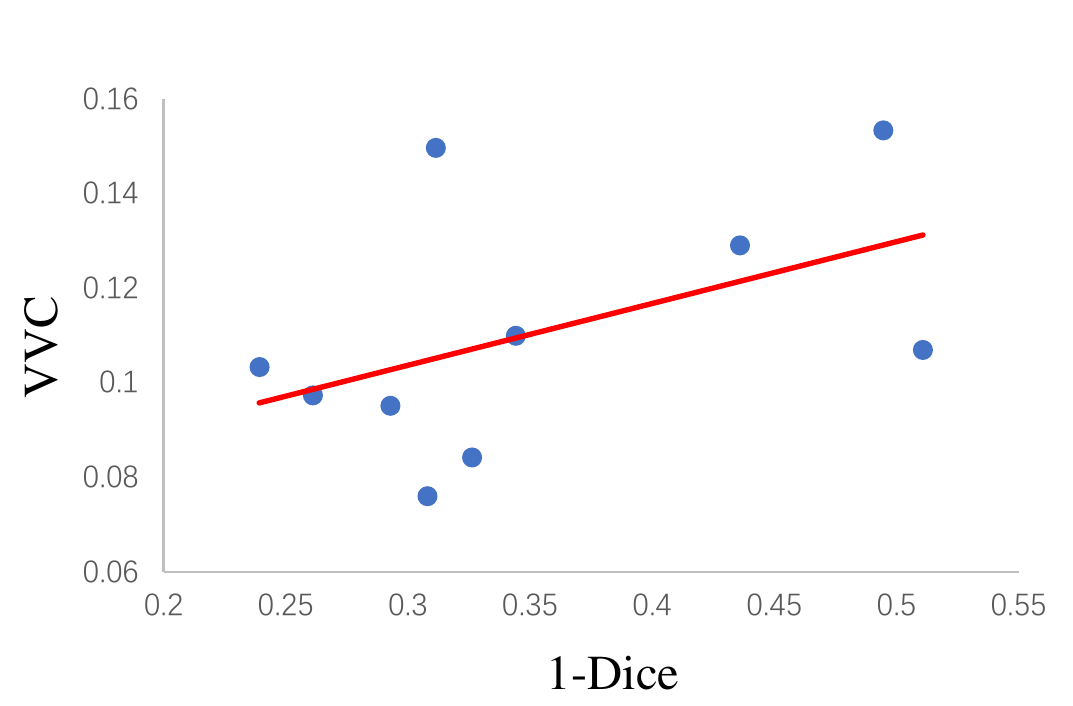}
	\caption{Correlation between structure-level uncertainty in terms of VVC and GTV segmentation error in terms of 1-Dice.}
	\label{fig:vvc}
\end{figure}
In order to study the use of uncertainty estimation methods to display erroneous segmentation, we first measured the uncertainty and segmentation error at pixel level. Since we used six models for ensemble in the binary segmentation task, we only obtained four possible uncertainty values. For each of four uncertainty levels, we calculated the error rate in segmentation results. Table 4 shows the numerical results of pixel-wise prediction error rate at each uncertainty level based on our test images. We chose a bounding box containing the entire head to calculate pixel-wise segmentation error rate. Since the bounding box contains most of the easily segmented background regions, the error rate corresponding to first uncertainty level is very low. Fig.~\ref{fig:pixel_error} shows the pixel-level prediction error rate at different uncertainty levels. It can be seen from Fig.~\ref{fig:pixel_error} that as the uncertainty increases, the segmentation error rate of the model also increases. Therefore, by measuring uncertainty of the model, it can effectively guide the model segmentation or increase the attention of doctors to the mis-segmentation areas.

\par We also measured the structure-level uncertainty in terms of VVC and investigated its relationship with 1-Dice. Fig.~\ref{fig:vvc} indicated the joint distribution of VVC and 1-Dice for different predictions of models. As shown in the figure, the value of VVC increases as the value of 1-Dice becomes larger. The comparison of VVC and 1-Dice shows that structure-wise uncertainty estimation is highly related to segmentation error. In other words, the higher the structure-wise uncertainty, the worse the segmentation accuracy. Similar to pixel-wise uncertainty, we can assess the quality of segmentation results based on structure-wise uncertainty, which can guide radiologists to focus on the images with poor segmentation to improve the efficiency of clinical decision making.

\begin{table}
\scalebox{0.9}{
\centering
\begin{tabular}{c c c c c}
\hline
\textbf{Patient} & level 1(\%) & level 2(\%) & level 3(\%) & level 4(\%)\\
\hline
    Patient1 & 0.21 & 23 & 44 & 69 \\
    Patient2 & 0.16 & 36 & 54 & 72 \\
    Patient3 & 0.50 & 36 & 41 & 57\\
    Patient4 & 0.21 & 28 & 55 & 86 \\
    Patient5 & 0.14 & 26 & 40 & 48 \\
    Patient6 & 0.33 & 28 & 50 & 76\\
    Patient7 & 0.22 & 36 & 54 & 56 \\
    Patient8 & 0.89 & 39 & 42 & 28 \\
    Patient9 & 0.57 & 45 & 52 & 60\\
    Patient10 & 0.93 & 34 & 29 & 20\\
    Average & 0.42 & 33 & 46 & 58 \\
\hline
\end{tabular}}
\caption{Pixel-level prediction error rate at different uncertainty levels. Four uncertainty values (0, 0.45, 0.63, 0.69) are separately denoted by level1, level2, level3, level4.}
\end{table}

\section{Discussion and Conclusion}
In this work, we propose a 2.5D CNN for GTV segmentation from CT images of NPC. Our network is designed for images with high in-plane resolution and low through-plane resolution. For this kind of data, our network shows better performance than 3D UNet~\cite{ronneberger2015u} and 3D VNet~\cite{milletari2016v}. For small target areas such as GTV, we propose a spatial attention mechanism and use it in conjunction with PE blocks~\cite{rickmann2019project}. The experimental results verify that our attention module can guide the network to better focus on the target area especially the boundary area when learning, which can improve the accuracy of segmentation. We also adopt multi-scale sampling for training that makes use of both local and global features for segmentation. By performing model ensemble with different scales of feature information, we can get more robust segmentation results. At the same time, we also estimate the uncertainty of models by model ensemble. Through the uncertainty analysis, we find that when performing GTV segmentation task, the uncertainty of GTV's boundaries or the low-contrast area which is difficult to segment is usually high. This also confirms why our AM pays more attention to the boundaries of GTV could enable our 2.5D CNN to obtain better segmentation results. As we mentioned before, the automatic segmentation for GTV of NPC task faces the challenge that NPC only takes up a small region in the whole head and neck CT image, which brings a large imbalance between the segmentation target and the background. We used Dice loss function \cite{milletari2016v} for training as it can effectively solve the problem of class imbalance and get good segmentation results. However, combining Dice loss with other common loss functions like Cross Entropy loss function can be considered for potentially higher performance \cite{wong20183d}. Afterwards, we further explore the relationship between uncertainty and error rate in both pixel-level and structure-level. Through the experimental results, we find that the higher the uncertainty, the higher the error rate, which means the high uncertainty region usually corresponds to a high chance to be mis-segmented. Due to this correlation between them, uncertainty is capable to play an important role in both guiding the improvement of segmentation results~\cite{wang2019aleatoric} and guiding the doctors to focus on specific areas during clinical diagnosis to improve the accuracy and efficiency of clinical decision, which will be explored in our future works. In addition, we will also consider some possibilities of using the anatomical structure information of neighboring organs for better segmentation~\cite{ren2018interleaved} in our future work.

\section*{Acknowledgments} 
This work was supported by the National Natural Science Foundations of China [81771921] funding.






\bibliographystyle{elsarticle-num-names}
\bibliography{sample}

\begin{thebibliography}{42}
\expandafter\ifx\csname natexlab\endcsname\relax\def\natexlab#1{#1}\fi
\providecommand{\url}[1]{\texttt{#1}}
\providecommand{\href}[2]{#2}
\providecommand{\path}[1]{#1}
\providecommand{\DOIprefix}{doi:}
\providecommand{\ArXivprefix}{arXiv:}
\providecommand{\URLprefix}{URL: }
\providecommand{\Pubmedprefix}{pmid:}
\providecommand{\doi}[1]{\href{http://dx.doi.org/#1}{\path{#1}}}
\providecommand{\Pubmed}[1]{\href{pmid:#1}{\path{#1}}}
\providecommand{\bibinfo}[2]{#2}
\ifx\xfnm\relax \def\xfnm[#1]{\unskip,\space#1}\fi
\bibitem[{Chang and Adami(2006)}]{chang2006enigmatic}
\bibinfo{author}{E.~T. Chang}, \bibinfo{author}{H.-O. Adami},
\newblock \bibinfo{title}{The enigmatic epidemiology of nasopharyngeal
  carcinoma},
\newblock \bibinfo{journal}{Cancer Epidemiology and Prevention Biomarkers}
  \bibinfo{volume}{15} (\bibinfo{year}{2006}) \bibinfo{pages}{1765--1777}.
\bibitem[{Xu et~al.(2015)Xu, Zhu, Hu, and Shen}]{xu2015omission}
\bibinfo{author}{T.~Xu}, \bibinfo{author}{G.~Zhu}, \bibinfo{author}{C.~Hu},
  \bibinfo{author}{C.~Shen},
\newblock \bibinfo{title}{Omission of chemotherapy in early-stage
  nasopharyngeal carcinoma treated with intensity modulated radiation therapy:
  A paired cohort study},
\newblock \bibinfo{journal}{International Journal of Radiation Oncology•
  Biology• Physics} \bibinfo{volume}{93} (\bibinfo{year}{2015})
  \bibinfo{pages}{E293--E294}.
\bibitem[{Long et~al.(2015)Long, Shelhamer, and Darrell}]{long2015fully}
\bibinfo{author}{J.~Long}, \bibinfo{author}{E.~Shelhamer},
  \bibinfo{author}{T.~Darrell},
\newblock \bibinfo{title}{Fully convolutional networks for semantic
  segmentation},
\newblock in: \bibinfo{booktitle}{Proceedings of the IEEE conference on
  computer vision and pattern recognition}, \bibinfo{year}{2015}, pp.
  \bibinfo{pages}{3431--3440}.
\bibitem[{Ronneberger et~al.(2015)Ronneberger, Fischer, and
  Brox}]{ronneberger2015u}
\bibinfo{author}{O.~Ronneberger}, \bibinfo{author}{P.~Fischer},
  \bibinfo{author}{T.~Brox},
\newblock \bibinfo{title}{U-net: Convolutional networks for biomedical image
  segmentation},
\newblock in: \bibinfo{booktitle}{International Conference on Medical image
  computing and computer-assisted intervention},
  \bibinfo{organization}{Springer}, \bibinfo{year}{2015}, pp.
  \bibinfo{pages}{234--241}.
\bibitem[{Milletari et~al.(2016)Milletari, Navab, and Ahmadi}]{milletari2016v}
\bibinfo{author}{F.~Milletari}, \bibinfo{author}{N.~Navab},
  \bibinfo{author}{S.-A. Ahmadi},
\newblock \bibinfo{title}{V-net: Fully convolutional neural networks for
  volumetric medical image segmentation},
\newblock in: \bibinfo{booktitle}{2016 Fourth International Conference on 3D
  Vision (3DV)}, \bibinfo{organization}{IEEE}, \bibinfo{year}{2016}, pp.
  \bibinfo{pages}{565--571}.
\bibitem[{Kamnitsas et~al.(2017)Kamnitsas, Ledig, Newcombe, Simpson, Kane,
  Menon, Rueckert, and Glocker}]{kamnitsas2017efficient}
\bibinfo{author}{K.~Kamnitsas}, \bibinfo{author}{C.~Ledig},
  \bibinfo{author}{V.~F. Newcombe}, \bibinfo{author}{J.~P. Simpson},
  \bibinfo{author}{A.~D. Kane}, \bibinfo{author}{D.~K. Menon},
  \bibinfo{author}{D.~Rueckert}, \bibinfo{author}{B.~Glocker},
\newblock \bibinfo{title}{Efficient multi-scale 3d cnn with fully connected crf
  for accurate brain lesion segmentation},
\newblock \bibinfo{journal}{Medical image analysis} \bibinfo{volume}{36}
  (\bibinfo{year}{2017}) \bibinfo{pages}{61--78}.
\bibitem[{Shi et~al.(2011)Shi, Zhuang, Wolz, Simon, Tung, Wang, Ourselin,
  Edwards, Razavi, and Rueckert}]{shi2011multi}
\bibinfo{author}{W.~Shi}, \bibinfo{author}{X.~Zhuang},
  \bibinfo{author}{R.~Wolz}, \bibinfo{author}{D.~Simon},
  \bibinfo{author}{K.~Tung}, \bibinfo{author}{H.~Wang},
  \bibinfo{author}{S.~Ourselin}, \bibinfo{author}{P.~Edwards},
  \bibinfo{author}{R.~Razavi}, \bibinfo{author}{D.~Rueckert},
\newblock \bibinfo{title}{A multi-image graph cut approach for cardiac image
  segmentation and uncertainty estimation},
\newblock in: \bibinfo{booktitle}{International Workshop on Statistical Atlases
  and Computational Models of the Heart}, \bibinfo{organization}{Springer},
  \bibinfo{year}{2011}, pp. \bibinfo{pages}{178--187}.
\bibitem[{Jungo et~al.(2018)Jungo, Meier, Ermis, Blatti-Moreno, Herrmann,
  Wiest, and Reyes}]{jungo2018effect}
\bibinfo{author}{A.~Jungo}, \bibinfo{author}{R.~Meier},
  \bibinfo{author}{E.~Ermis}, \bibinfo{author}{M.~Blatti-Moreno},
  \bibinfo{author}{E.~Herrmann}, \bibinfo{author}{R.~Wiest},
  \bibinfo{author}{M.~Reyes},
\newblock \bibinfo{title}{On the effect of inter-observer variability for a
  reliable estimation of uncertainty of medical image segmentation},
\newblock in: \bibinfo{booktitle}{International Conference on Medical Image
  Computing and Computer-Assisted Intervention},
  \bibinfo{organization}{Springer}, \bibinfo{year}{2018}, pp.
  \bibinfo{pages}{682--690}.
\bibitem[{Lee et~al.(2005)Lee, Yeung, King, Leung, and
  Ahuja}]{lee2005segmentation}
\bibinfo{author}{F.~K. Lee}, \bibinfo{author}{D.~K. Yeung},
  \bibinfo{author}{A.~D. King}, \bibinfo{author}{S.~Leung},
  \bibinfo{author}{A.~Ahuja},
\newblock \bibinfo{title}{Segmentation of nasopharyngeal carcinoma (npc)
  lesions in mr images},
\newblock \bibinfo{journal}{International Journal of Radiation Oncology*
  Biology* Physics} \bibinfo{volume}{61} (\bibinfo{year}{2005})
  \bibinfo{pages}{608--620}.
\bibitem[{Han et~al.(2008)Han, Hoogeman, Levendag, Hibbard, Teguh, Voet, Cowen,
  and Wolf}]{han2008atlas}
\bibinfo{author}{X.~Han}, \bibinfo{author}{M.~S. Hoogeman},
  \bibinfo{author}{P.~C. Levendag}, \bibinfo{author}{L.~S. Hibbard},
  \bibinfo{author}{D.~N. Teguh}, \bibinfo{author}{P.~Voet},
  \bibinfo{author}{A.~C. Cowen}, \bibinfo{author}{T.~K. Wolf},
\newblock \bibinfo{title}{Atlas-based auto-segmentation of head and neck ct
  images},
\newblock in: \bibinfo{booktitle}{International Conference on Medical Image
  Computing and Computer-assisted Intervention},
  \bibinfo{organization}{Springer}, \bibinfo{year}{2008}, pp.
  \bibinfo{pages}{434--441}.
\bibitem[{Berthon et~al.(2016)Berthon, Marshall, Evans, and
  Spezi}]{berthon2016atlaas}
\bibinfo{author}{B.~Berthon}, \bibinfo{author}{C.~Marshall},
  \bibinfo{author}{M.~Evans}, \bibinfo{author}{E.~Spezi},
\newblock \bibinfo{title}{Atlaas: an automatic decision tree-based learning
  algorithm for advanced image segmentation in positron emission tomography},
\newblock \bibinfo{journal}{Physics in Medicine \& Biology}
  \bibinfo{volume}{61} (\bibinfo{year}{2016}) \bibinfo{pages}{4855}.
\bibitem[{Berthon et~al.(2017)Berthon, Evans, Marshall, Palaniappan, Cole,
  Jayaprakasam, Rackley, and Spezi}]{berthon2017head}
\bibinfo{author}{B.~Berthon}, \bibinfo{author}{M.~Evans},
  \bibinfo{author}{C.~Marshall}, \bibinfo{author}{N.~Palaniappan},
  \bibinfo{author}{N.~Cole}, \bibinfo{author}{V.~Jayaprakasam},
  \bibinfo{author}{T.~Rackley}, \bibinfo{author}{E.~Spezi},
\newblock \bibinfo{title}{Head and neck target delineation using a novel pet
  automatic segmentation algorithm},
\newblock \bibinfo{journal}{Radiotherapy and Oncology} \bibinfo{volume}{122}
  (\bibinfo{year}{2017}) \bibinfo{pages}{242--247}.
\bibitem[{Mohammed et~al.(2017)Mohammed, Ghani, Hamed, Ibrahim, and
  Abdullah}]{mohammed2017artificial}
\bibinfo{author}{M.~A. Mohammed}, \bibinfo{author}{M.~K.~A. Ghani},
  \bibinfo{author}{R.~I. Hamed}, \bibinfo{author}{D.~A. Ibrahim},
  \bibinfo{author}{M.~K. Abdullah},
\newblock \bibinfo{title}{Artificial neural networks for automatic segmentation
  and identification of nasopharyngeal carcinoma},
\newblock \bibinfo{journal}{Journal of computational science}
  \bibinfo{volume}{21} (\bibinfo{year}{2017}) \bibinfo{pages}{263--274}.
\bibitem[{Men et~al.(2017)Men, Chen, Zhang, Zhang, Dai, Yi, and
  Li}]{men2017deep}
\bibinfo{author}{K.~Men}, \bibinfo{author}{X.~Chen},
  \bibinfo{author}{Y.~Zhang}, \bibinfo{author}{T.~Zhang},
  \bibinfo{author}{J.~Dai}, \bibinfo{author}{J.~Yi}, \bibinfo{author}{Y.~Li},
\newblock \bibinfo{title}{Deep deconvolutional neural network for target
  segmentation of nasopharyngeal cancer in planning computed tomography
  images},
\newblock \bibinfo{journal}{Frontiers in oncology} \bibinfo{volume}{7}
  (\bibinfo{year}{2017}) \bibinfo{pages}{315}.
\bibitem[{Ma et~al.(2017)Ma, Wu, and Zhou}]{ma2017automatic}
\bibinfo{author}{Z.~Ma}, \bibinfo{author}{X.~Wu}, \bibinfo{author}{J.~Zhou},
\newblock \bibinfo{title}{Automatic nasopharyngeal carcinoma segmentation in mr
  images with convolutional neural networks},
\newblock in: \bibinfo{booktitle}{2017 International Conference on the
  Frontiers and Advances in Data Science (FADS)}, \bibinfo{organization}{IEEE},
  \bibinfo{year}{2017}, pp. \bibinfo{pages}{147--150}.
\bibitem[{Guo et~al.(2019)Guo, Guo, Gong, Li et~al.}]{guo2019gross}
\bibinfo{author}{Z.~Guo}, \bibinfo{author}{N.~Guo}, \bibinfo{author}{K.~Gong},
  \bibinfo{author}{Q.~Li}, et~al.,
\newblock \bibinfo{title}{Gross tumor volume segmentation for head and neck
  cancer radiotherapy using deep dense multi-modality network},
\newblock \bibinfo{journal}{Physics in Medicine \& Biology}
  \bibinfo{volume}{64} (\bibinfo{year}{2019}) \bibinfo{pages}{205015}.
\bibitem[{Chen et~al.(2018)Chen, Qi, Yin, Li, Gong, and Wang}]{chen2018mmfnet}
\bibinfo{author}{H.~Chen}, \bibinfo{author}{Y.~Qi}, \bibinfo{author}{Y.~Yin},
  \bibinfo{author}{T.~Li}, \bibinfo{author}{G.~Gong},
  \bibinfo{author}{L.~Wang},
\newblock \bibinfo{title}{Mmfnet: A multi-modality mri fusion network for
  segmentation of nasopharyngeal carcinoma},
\newblock \bibinfo{journal}{arXiv preprint arXiv:1812.10033}
  (\bibinfo{year}{2018}).
\bibitem[{Oktay et~al.(2018)Oktay, Schlemper, Folgoc, Lee, Heinrich, Misawa,
  Mori, McDonagh, Hammerla, Kainz et~al.}]{oktay2018attention}
\bibinfo{author}{O.~Oktay}, \bibinfo{author}{J.~Schlemper},
  \bibinfo{author}{L.~L. Folgoc}, \bibinfo{author}{M.~Lee},
  \bibinfo{author}{M.~Heinrich}, \bibinfo{author}{K.~Misawa},
  \bibinfo{author}{K.~Mori}, \bibinfo{author}{S.~McDonagh},
  \bibinfo{author}{N.~Y. Hammerla}, \bibinfo{author}{B.~Kainz}, et~al.,
\newblock \bibinfo{title}{Attention u-net: Learning where to look for the
  pancreas},
\newblock \bibinfo{journal}{arXiv preprint arXiv:1804.03999}
  (\bibinfo{year}{2018}).
\bibitem[{Hu et~al.(2018)Hu, Shen, and Sun}]{hu2018squeeze}
\bibinfo{author}{J.~Hu}, \bibinfo{author}{L.~Shen}, \bibinfo{author}{G.~Sun},
\newblock \bibinfo{title}{Squeeze-and-excitation networks},
\newblock in: \bibinfo{booktitle}{Proceedings of the IEEE conference on
  computer vision and pattern recognition}, \bibinfo{year}{2018}, pp.
  \bibinfo{pages}{7132--7141}.
\bibitem[{Woo et~al.(2018)Woo, Park, Lee, and So~Kweon}]{woo2018cbam}
\bibinfo{author}{S.~Woo}, \bibinfo{author}{J.~Park}, \bibinfo{author}{J.-Y.
  Lee}, \bibinfo{author}{I.~So~Kweon},
\newblock \bibinfo{title}{Cbam: Convolutional block attention module},
\newblock in: \bibinfo{booktitle}{Proceedings of the European Conference on
  Computer Vision (ECCV)}, \bibinfo{year}{2018}, pp. \bibinfo{pages}{3--19}.
\bibitem[{Roy et~al.(2018)Roy, Navab, and Wachinger}]{roy2018concurrent}
\bibinfo{author}{A.~G. Roy}, \bibinfo{author}{N.~Navab},
  \bibinfo{author}{C.~Wachinger},
\newblock \bibinfo{title}{Concurrent spatial and channel ‘squeeze \&
  excitation’in fully convolutional networks},
\newblock in: \bibinfo{booktitle}{International Conference on Medical Image
  Computing and Computer-Assisted Intervention},
  \bibinfo{organization}{Springer}, \bibinfo{year}{2018}, pp.
  \bibinfo{pages}{421--429}.
\bibitem[{Wang et~al.(2018)Wang, Girshick, Gupta, and He}]{wang2018non}
\bibinfo{author}{X.~Wang}, \bibinfo{author}{R.~Girshick},
  \bibinfo{author}{A.~Gupta}, \bibinfo{author}{K.~He},
\newblock \bibinfo{title}{Non-local neural networks},
\newblock in: \bibinfo{booktitle}{Proceedings of the IEEE conference on
  computer vision and pattern recognition}, \bibinfo{year}{2018}, pp.
  \bibinfo{pages}{7794--7803}.
\bibitem[{Fu et~al.(2019)Fu, Liu, Tian, Li, Bao, Fang, and Lu}]{fu2019dual}
\bibinfo{author}{J.~Fu}, \bibinfo{author}{J.~Liu}, \bibinfo{author}{H.~Tian},
  \bibinfo{author}{Y.~Li}, \bibinfo{author}{Y.~Bao}, \bibinfo{author}{Z.~Fang},
  \bibinfo{author}{H.~Lu},
\newblock \bibinfo{title}{Dual attention network for scene segmentation},
\newblock in: \bibinfo{booktitle}{Proceedings of the IEEE Conference on
  Computer Vision and Pattern Recognition}, \bibinfo{year}{2019}, pp.
  \bibinfo{pages}{3146--3154}.
\bibitem[{Li et~al.(2018)Li, Xiong, An, and Wang}]{li2018pyramid}
\bibinfo{author}{H.~Li}, \bibinfo{author}{P.~Xiong}, \bibinfo{author}{J.~An},
  \bibinfo{author}{L.~Wang},
\newblock \bibinfo{title}{Pyramid attention network for semantic segmentation},
\newblock \bibinfo{journal}{arXiv preprint arXiv:1805.10180}
  (\bibinfo{year}{2018}).
\bibitem[{Chen et~al.(2016)Chen, Yang, Wang, Xu, and
  Yuille}]{chen2016attention}
\bibinfo{author}{L.-C. Chen}, \bibinfo{author}{Y.~Yang},
  \bibinfo{author}{J.~Wang}, \bibinfo{author}{W.~Xu}, \bibinfo{author}{A.~L.
  Yuille},
\newblock \bibinfo{title}{Attention to scale: Scale-aware semantic image
  segmentation},
\newblock in: \bibinfo{booktitle}{Proceedings of the IEEE conference on
  computer vision and pattern recognition}, \bibinfo{year}{2016}, pp.
  \bibinfo{pages}{3640--3649}.
\bibitem[{Huang et~al.(2019)Huang, Wang, Huang, Huang, Wei, and
  Liu}]{huang2019ccnet}
\bibinfo{author}{Z.~Huang}, \bibinfo{author}{X.~Wang},
  \bibinfo{author}{L.~Huang}, \bibinfo{author}{C.~Huang},
  \bibinfo{author}{Y.~Wei}, \bibinfo{author}{W.~Liu},
\newblock \bibinfo{title}{Ccnet: Criss-cross attention for semantic
  segmentation},
\newblock in: \bibinfo{booktitle}{Proceedings of the IEEE International
  Conference on Computer Vision}, \bibinfo{year}{2019}, pp.
  \bibinfo{pages}{603--612}.
\bibitem[{Li et~al.(2018)Li, Wu, Peng, Ernst, and Fu}]{li2018tell}
\bibinfo{author}{K.~Li}, \bibinfo{author}{Z.~Wu}, \bibinfo{author}{K.-C. Peng},
  \bibinfo{author}{J.~Ernst}, \bibinfo{author}{Y.~Fu},
\newblock \bibinfo{title}{Tell me where to look: Guided attention inference
  network},
\newblock in: \bibinfo{booktitle}{Proceedings of the IEEE Conference on
  Computer Vision and Pattern Recognition}, \bibinfo{year}{2018}, pp.
  \bibinfo{pages}{9215--9223}.
\bibitem[{Rickmann et~al.(2019)Rickmann, Sarasua, Roy, Navab, and
  Wachinger}]{rickmann2019project}
\bibinfo{author}{A.-M. Rickmann}, \bibinfo{author}{I.~Sarasua},
  \bibinfo{author}{A.~G. Roy}, \bibinfo{author}{N.~Navab},
  \bibinfo{author}{C.~Wachinger},
\newblock \bibinfo{title}{`project \& excite' modules for segmentation of
  volumetric medical scans},
\newblock in: \bibinfo{booktitle}{International Conference on Medical Image
  Computing and Computer-Assisted Intervention},
  \bibinfo{organization}{Springer}, \bibinfo{year}{2019}.
\bibitem[{Saad et~al.(2010)Saad, Hamarneh, and Moller}]{saad2010exploration}
\bibinfo{author}{A.~Saad}, \bibinfo{author}{G.~Hamarneh},
  \bibinfo{author}{T.~Moller},
\newblock \bibinfo{title}{Exploration and visualization of segmentation
  uncertainty using shape and appearance prior information},
\newblock \bibinfo{journal}{IEEE Transactions on Visualization and Computer
  Graphics} \bibinfo{volume}{16} (\bibinfo{year}{2010})
  \bibinfo{pages}{1366--1375}.
\bibitem[{Prassni et~al.(2010)Prassni, Ropinski, and
  Hinrichs}]{prassni2010uncertainty}
\bibinfo{author}{J.-S. Prassni}, \bibinfo{author}{T.~Ropinski},
  \bibinfo{author}{K.~Hinrichs},
\newblock \bibinfo{title}{Uncertainty-aware guided volume segmentation},
\newblock \bibinfo{journal}{IEEE transactions on visualization and computer
  graphics} \bibinfo{volume}{16} (\bibinfo{year}{2010})
  \bibinfo{pages}{1358--1365}.
\bibitem[{Sankaran et~al.(2015)Sankaran, Grady, and Taylor}]{sankaran2015fast}
\bibinfo{author}{S.~Sankaran}, \bibinfo{author}{L.~Grady},
  \bibinfo{author}{C.~A. Taylor},
\newblock \bibinfo{title}{Fast computation of hemodynamic sensitivity to lumen
  segmentation uncertainty},
\newblock \bibinfo{journal}{IEEE transactions on medical imaging}
  \bibinfo{volume}{34} (\bibinfo{year}{2015}) \bibinfo{pages}{2562--2571}.
\bibitem[{Li et~al.(2017)Li, Wang, Fidon, Ourselin, Cardoso, and
  Vercauteren}]{li2017compactness}
\bibinfo{author}{W.~Li}, \bibinfo{author}{G.~Wang}, \bibinfo{author}{L.~Fidon},
  \bibinfo{author}{S.~Ourselin}, \bibinfo{author}{M.~J. Cardoso},
  \bibinfo{author}{T.~Vercauteren},
\newblock \bibinfo{title}{On the compactness, efficiency, and representation of
  3d convolutional networks: brain parcellation as a pretext task},
\newblock in: \bibinfo{booktitle}{International Conference on Information
  Processing in Medical Imaging}, \bibinfo{organization}{Springer},
  \bibinfo{year}{2017}, pp. \bibinfo{pages}{348--360}.
\bibitem[{Zhu and Zabaras(2018)}]{zhu2018bayesian}
\bibinfo{author}{Y.~Zhu}, \bibinfo{author}{N.~Zabaras},
\newblock \bibinfo{title}{Bayesian deep convolutional encoder--decoder networks
  for surrogate modeling and uncertainty quantification},
\newblock \bibinfo{journal}{Journal of Computational Physics}
  \bibinfo{volume}{366} (\bibinfo{year}{2018}) \bibinfo{pages}{415--447}.
\bibitem[{Lakshminarayanan et~al.(2017)Lakshminarayanan, Pritzel, and
  Blundell}]{lakshminarayanan2017simple}
\bibinfo{author}{B.~Lakshminarayanan}, \bibinfo{author}{A.~Pritzel},
  \bibinfo{author}{C.~Blundell},
\newblock \bibinfo{title}{Simple and scalable predictive uncertainty estimation
  using deep ensembles},
\newblock in: \bibinfo{booktitle}{Advances in Neural Information Processing
  Systems}, \bibinfo{year}{2017}, pp. \bibinfo{pages}{6402--6413}.
\bibitem[{Kendall and Gal(2017)}]{kendall2017uncertainties}
\bibinfo{author}{A.~Kendall}, \bibinfo{author}{Y.~Gal},
\newblock \bibinfo{title}{What uncertainties do we need in bayesian deep
  learning for computer vision?},
\newblock in: \bibinfo{booktitle}{Advances in neural information processing
  systems}, \bibinfo{year}{2017}, pp. \bibinfo{pages}{5574--5584}.
\bibitem[{Wang et~al.(2019)Wang, Li, Aertsen, Deprest, Ourselin, and
  Vercauteren}]{wang2019aleatoric}
\bibinfo{author}{G.~Wang}, \bibinfo{author}{W.~Li},
  \bibinfo{author}{M.~Aertsen}, \bibinfo{author}{J.~Deprest},
  \bibinfo{author}{S.~Ourselin}, \bibinfo{author}{T.~Vercauteren},
\newblock \bibinfo{title}{Aleatoric uncertainty estimation with test-time
  augmentation for medical image segmentation with convolutional neural
  networks},
\newblock \bibinfo{journal}{Neurocomputing} \bibinfo{volume}{338}
  (\bibinfo{year}{2019}) \bibinfo{pages}{34--45}.
\bibitem[{Vu et~al.(2019)Vu, Grimbergen, Nyholm, and
  L{\"o}fstedt}]{vu2019evaluation}
\bibinfo{author}{M.~H. Vu}, \bibinfo{author}{G.~Grimbergen},
  \bibinfo{author}{T.~Nyholm}, \bibinfo{author}{T.~L{\"o}fstedt},
\newblock \bibinfo{title}{Evaluation of multi-slice inputs to convolutional
  neural networks for medical image segmentation},
\newblock \bibinfo{journal}{arXiv preprint arXiv:1912.09287}
  (\bibinfo{year}{2019}).
\bibitem[{Prasoon et~al.(2013)Prasoon, Petersen, Igel, Lauze, Dam, and
  Nielsen}]{prasoon2013deep}
\bibinfo{author}{A.~Prasoon}, \bibinfo{author}{K.~Petersen},
  \bibinfo{author}{C.~Igel}, \bibinfo{author}{F.~Lauze},
  \bibinfo{author}{E.~Dam}, \bibinfo{author}{M.~Nielsen},
\newblock \bibinfo{title}{Deep feature learning for knee cartilage segmentation
  using a triplanar convolutional neural network},
\newblock in: \bibinfo{booktitle}{International conference on medical image
  computing and computer-assisted intervention},
  \bibinfo{organization}{Springer}, \bibinfo{year}{2013}, pp.
  \bibinfo{pages}{246--253}.
\bibitem[{Zhu et~al.(2019)Zhu, Huang, Zeng, Chen, Liu, Qian, Du, Fan, and
  Xie}]{zhu2019anatomynet}
\bibinfo{author}{W.~Zhu}, \bibinfo{author}{Y.~Huang},
  \bibinfo{author}{L.~Zeng}, \bibinfo{author}{X.~Chen},
  \bibinfo{author}{Y.~Liu}, \bibinfo{author}{Z.~Qian}, \bibinfo{author}{N.~Du},
  \bibinfo{author}{W.~Fan}, \bibinfo{author}{X.~Xie},
\newblock \bibinfo{title}{Anatomynet: Deep learning for fast and fully
  automated whole-volume segmentation of head and neck anatomy},
\newblock \bibinfo{journal}{Medical physics} \bibinfo{volume}{46}
  (\bibinfo{year}{2019}) \bibinfo{pages}{576--589}.
\bibitem[{Bakas et~al.(2018)Bakas, Reyes, Jakab, Bauer, Rempfler, Crimi,
  Shinohara, Berger, Ha, Rozycki et~al.}]{bakas2018identifying}
\bibinfo{author}{S.~Bakas}, \bibinfo{author}{M.~Reyes},
  \bibinfo{author}{A.~Jakab}, \bibinfo{author}{S.~Bauer},
  \bibinfo{author}{M.~Rempfler}, \bibinfo{author}{A.~Crimi},
  \bibinfo{author}{R.~T. Shinohara}, \bibinfo{author}{C.~Berger},
  \bibinfo{author}{S.~M. Ha}, \bibinfo{author}{M.~Rozycki}, et~al.,
\newblock \bibinfo{title}{Identifying the best machine learning algorithms for
  brain tumor segmentation, progression assessment, and overall survival
  prediction in the brats challenge},
\newblock \bibinfo{journal}{arXiv preprint arXiv:1811.02629}
  (\bibinfo{year}{2018}).
\bibitem[{Wong et~al.(2018)Wong, Moradi, Tang, and Syeda-Mahmood}]{wong20183d}
\bibinfo{author}{K.~C. Wong}, \bibinfo{author}{M.~Moradi},
  \bibinfo{author}{H.~Tang}, \bibinfo{author}{T.~Syeda-Mahmood},
\newblock \bibinfo{title}{3d segmentation with exponential logarithmic loss for
  highly unbalanced object sizes},
\newblock in: \bibinfo{booktitle}{International Conference on Medical Image
  Computing and Computer-Assisted Intervention},
  \bibinfo{organization}{Springer}, \bibinfo{year}{2018}, pp.
  \bibinfo{pages}{612--619}.
\bibitem[{Ren et~al.(2018)Ren, Xiang, Nie, Shao, Zhang, Shen, and
  Wang}]{ren2018interleaved}
\bibinfo{author}{X.~Ren}, \bibinfo{author}{L.~Xiang}, \bibinfo{author}{D.~Nie},
  \bibinfo{author}{Y.~Shao}, \bibinfo{author}{H.~Zhang},
  \bibinfo{author}{D.~Shen}, \bibinfo{author}{Q.~Wang},
\newblock \bibinfo{title}{Interleaved 3d-cnn s for joint segmentation of
  small-volume structures in head and neck ct images},
\newblock \bibinfo{journal}{Medical physics} \bibinfo{volume}{45}
  (\bibinfo{year}{2018}) \bibinfo{pages}{2063--2075}.

\end{thebibliography}







\end{document}